\documentclass[11pt,letterpaper]{article}

\usepackage[margin=1in]{geometry}
\usepackage[T1]{fontenc}
\usepackage{lmodern}
\usepackage[english]{babel}
\usepackage{microtype}

\usepackage{amsmath,amssymb,amsthm}
\usepackage{mathtools}
\usepackage{bm}
\numberwithin{equation}{section}

\usepackage{graphicx}
\graphicspath{{figures/}}
\usepackage{booktabs}
\usepackage{siunitx}
\usepackage{float}

\usepackage{xcolor}
\usepackage{listings}
\lstdefinestyle{networkpython}{
    language=Python,
    basicstyle=\fontsize{9}{10.5}\selectfont\ttfamily,
    backgroundcolor=\color{black!5},
    keywordstyle=\color{blue!65!black}\bfseries,
    morekeywords=[2]{Network,Distributor,NetworkBasis,Field,LBVP,
        add_equation,add_vertex_constraint,build_solver,solve,
        local_grid,local_physical_grid},
    keywordstyle=[2]\color{blue!65!black},
    commentstyle=\color{green!35!black},
    stringstyle=\color{red!60!black},
    frame=single,
    rulecolor=\color{black!20},
    framerule=0.4pt,
    framesep=6pt,
    xleftmargin=6pt,
    xrightmargin=6pt,
    columns=fullflexible,
    keepspaces=true,
    showstringspaces=false,
    upquote=true,
    breaklines=true,
    breakatwhitespace=true,
    tabsize=4
}

\usepackage[numbers,sort&compress]{natbib}

\usepackage[hidelinks]{hyperref}
\usepackage[nameinlink,noabbrev]{cleveref}

\newcommand{\grad}{\operatorname{grad}}
\renewcommand{\div}{\operatorname{div}}

\newcommand{\dat}{\!\cdot\!}
\newcommand{\tr}{\operatorname{tr}}

\title{A scalable computational framework for PDEs on physical networks}
\author{Sidney Holden$^{1}$, Keaton J. Burns$^{2,3}$, Daniel Lecoanet$^{3}$,\\ Jeffrey S. Oishi$^{4}$, and Benjamin P. Brown$^{5}$, Geoffrey M. Vasil$^{6}$\\[0.6em]
\parbox{0.95\textwidth}{\centering\small
$^{1}$Center for Computational Biology, Flatiron Institute, New York, NY, USA\\[0.25em]
$^{2}$Department of Mathematics, Massachusetts Institute of Technology, Cambridge, MA, USA\\[0.25em]
$^{3}$Center for Computational Astrophysics, Flatiron Institute, New York, NY, USA\\[0.25em]
$^{4}$Program in Integrated Applied Mathematics and Department of Mechanical Engineering, University of New Hampshire, Durham, NH, USA\\[0.25em]
$^{5}$Department of Astrophysical and Planetary Sciences and\\
Laboratory for Atmospheric and Space Physics, University of Colorado Boulder, Boulder, CO, USA \\ 
[0.25em]
$^{6}$School of Mathematics and Maxwell Institute for Mathematical Sciences,\\
University of Edinburgh, Edinburgh, UK
}}
\date{\today}

\begin{document}

\maketitle

\begin{abstract}

Transport, reaction and wave propagation on physical networks depend on dynamics along edges and interactions at junctions. We present a flexible framework for modeling physical networks at scale. A network calculus expresses edge dynamics and vertex contributions as fields over the whole network. Differential operators satisfy integration by parts-identities that connect local dynamics to global conservation laws, and explicit vertex constraints allow a variety of physical junction conditions. We implement this formalism using the open-source Dedalus spectral framework. This accommodates heterogeneous edge lengths, material properties and spectral resolutions. Initial-value, boundary-value and eigenvalue problems share a parallel solution method that eliminates internal unknowns independently on subgraphs and couples their solutions through shared boundary vertex interfaces. We verify numerical convergence and demonstrate strong and weak scaling on networks with millions of edges. We present applications to localized optical resonances in photonic networks, preferential channel formation during mineral replacement in evolving pore networks, and nonlinear dispersive waves for KdV on a periodic network. 

\end{abstract}

\section{Introduction}
\label{sec:network-equations}

In many applications, the edges of a network represent physical connections with distance along their length. 
Notable examples include electrical cables, railway lines and blood vessels. 
Such physical networks, or \emph{metric graphs}, support differential equations along their edges, coupled by conditions at the vertices. 

In this paper we develop a calculus for formulating and solving differential equations on networks as one would on an ordinary continuous domain, such as a torus or disk. 
Abstract fields describe material properties and physical variables, while differential operators encode both local derivatives and vertex contributions. 
The gradient and divergence satisfy integration-by-parts identities that connect edge dynamics to conservation across the network. 
They also satisfy standard calculus rules such as the Leibniz identity. 

Our numerical implementation is built on the Dedalus Project, an open-source framework for solving differential equations using spectral methods \citep{burns_etal_2020}. 
Earlier developments extended this framework to geometries such as disks and balls \citep{vasil_etal_2016,Vasil2019,Lecoanet2019}. 
The project combines mathematical structure, flexible model specification and robust numerical algorithms within a shared implementation.  
Shared algorithms combine testing and maintenance, allowing improvements to benefit many applications. 
Documented methods and accessible simulation and analysis scripts support scrutiny, reuse and reproducibility \citep{Oishi2018}. 
Recent large-scale datasets illustrate the growing scale of whole-network reconstructions: MICrONS maps approximately $5.2\times10^8$
synapses in a mouse visual cortex~\cite{MICrONS2025}, Ji et al. reconstructs approximately 6.3 million blood vessel branches per mouse brain \cite{Ji2021}, and NASA/CNES SWOT provides river geometry and water-storage observations for 126,674 river reaches worldwide, including 21,123 reaches in the Amazon basin \cite{Cerbelaud2026}. 
We aim for our infrastructure to enable new network science in a variety of contexts at scale.

We construct a network basis from Dedalus' cardinal basis for finite sets.
Discrete edge, vertex and endpoint labels enter field representations combined with continuous coordinates.
Each edge has its own spectral expansion, and the graph's connectivity is used to prescribe the conditions at vertices.
Each physical variable is represented as a single network field, accommodating different edge lengths and material properties.
The separation between edge dynamics and vertex coupling enables parallel computation over subgraphs.
We partition the network, eliminate internal unknowns independently on each subgraph, and solve a reduced system coupling the interfaces.
Partitions balance computational work and reduce communication.
The subgraph distribution introduces no additional discretisation approximation.

\subsection{Related work}

\citet{Berkolaiko2012} provide the primary reference for the mathematical setting used here, including metric graphs, differential operators, vertex conditions and spectral theory. Earlier surveys describe the basic operator structures and their connections to mathematical physics \citep{Kuchment2004,Kuchment2008}. 
A recent computational treatment discusses dynamical processes on metric graphs \citep{Bttcher2025}.

Early analytical work establishes spectral characterisations and solvability for network systems \citep{vonBelow1985,vonBelow1988}. 
Operator formulations classify self-adjoint and skew-adjoint interactions \citep{Kostrykin1999,Schubert2015} and extend the analysis to non-self-adjoint problems \citep{Hussein2014}. 
Limits of diffusion processes and thin domains provide another route to graph models, deriving effective operators and junction conditions from continuous geometries \citep{Freidlin1993,Exner2005}.

Variational and semigroup methods organise diffusion and wave equations within common frameworks \citep{Fijavz2007,Mugnolo2014}. 
Dynamic vertex conditions couple edge equations to additional evolution at the junctions \citep{Mugnolo2006,KramarFijav2021b}. 
These formulations accommodate concentrated storage and other vertex dynamics, as in the heat equation below. 
General hyperbolic and Hamiltonian systems connect well-posedness to the structure of their energy balances \citep{KramarFijav2021a,Banasiak2022}. 
Higher-order network operators require corresponding extensions of the admissible fields and junction data \citep{Gregorio2020}.

Nonlinear dispersive equations form another substantial part of the literature. 
Studies of nonlinear Schr\"{o}dinger equations address solitons, standing waves and stability \citep{Noja2014,Kairzhan2022}, including the dependence of ground-state existence on graph topology \citep{Adami2014}. 
Airy and Korteweg-de Vries (KdV) equations make the relationship between edge orientation, vertex coupling and well-posedness particularly explicit \citep{Mugnolo2018,Cavalcante2018,Angulo2024}. 
Related models include travelling waves for the Benjamin--Bona--Mahony equation on networks \citep{Mugnolo2014a}.

Many analytical investigations include computational examples that resolve solution branches and examine stability. 
Nonlinear Schr\"{o}dinger equations show how loops and junctions affect bifurcations and localisation \citep{Noja2014,Goodman2019,Kairzhan2021}. Physical applications include vascular flow \citep{Sherwin2003}, gas pipelines and compressors \citep{KBanda2006,Herty2007}, road traffic \citep{Coclite2005}, reactive transport in catalytic networks \citep{Wallace2017}, and network lasers \citep{Saxena2022}. These examples combine differential equations along edges with junction relations determined by the physical model.

Compatible discretisations preserve relations between fields and differential operators. Finite element exterior calculus connects these relations to topology and numerical stability \citep{Arnold2010}, while mimetic finite differences preserve integral and adjoint identities \citep{Hyman1999}. Recent examples include twisted and Hessian complexes on polygonal meshes \citep{DiPietro2025}, distributional Hessian and div-div complexes containing polynomial fields and Dirac measures \citep{Hu2025}, and finite-element calculations of spectra and pseudospectra for advection--diffusion of differential forms \citep{Boffi2026}. Section~\ref{sec:compatible-spectral} gives a simple polynomial example and relates these ideas to spectral discretisation.

Numerical methods address both approximation along individual edges and coupling across the network. Finite-element formulations provide discretisation and error analysis for graph differential operators \citep{Arioli2017}. Comparisons of spectral, finite-difference and discontinuous Galerkin methods demonstrate different treatments of vertex conditions in eigenvalue, Poisson and wave problems \citep{Brio2022}. Further developments include fast graph Fourier transforms \citep{Carlson2026}, domain decomposition for elliptic equations \citep{Kovcs2025}, and model reduction that preserves conservation and dissipation properties \citep{Egger2018}. Other methods treat fractional elliptic operators \citep{Bolin2023} and compute nonlinear stationary states through gradient flows \citep{Besse2022a}.

Several software packages provide tools for metric-graph problems. GraFiDi \citep{Besse2022a} supplies finite-difference operators in Python, with methods for computing ground states and time evolution of nonlinear Schr\"{o}dinger equations. QGLAB \citep{Goodman2025} combines finite-difference and Chebyshev discretisations in MATLAB, supporting Poisson problems, eigenvalue calculations, linear and nonlinear time evolution, and continuation of steady solutions through bifurcations. The package netSALT \citep{Saxena2022} models photonic networks using physically realistic parameters. PETSc DMNetwork \citep{Abhyankar2020} provides an environment for parallel multiphysics simulation on networks, including topology and data management, partitioning and distribution, and access to PETSc's linear, nonlinear and time-integration solvers.

These packages provide precedents for separating graph construction and numerical operators from individual calculations. 
The present work develops a common calculus of graph fields, differential operators and junction balances, building upon the general Dedalus spectral framework.

\subsection{An illustrative example}

Heat diffusion on a three-edged star graph provides a simple example. The temperature on each edge is
\begin{align}
    T_{e}(t,x_{e}),
    \qquad
    x_{e}\in[0,\infty),
    \qquad
    e=1,\,2,\,3.
\end{align}
The edges meet at $x_{e}=0$, where the temperature is continuous,
\begin{align}
    T_{1}(t,0)=T_{2}(t,0)=T_{3}(t,0)=T_{0}(t),
\end{align}
where $T_{0}(t)$ is the common value at the central vertex. 
Each edge has a heat capacity per unit length, $C_{e}(x_{e})$, and conductivity, $K_{e}(x_{e})$. 
The edge equation is
\begin{align}
    C_{e}\,\partial_{t}T_{e}
    -\partial_{x_{e}}\!\left(K_{e}\,\partial_{x_{e}}T_{e}\right)
    \,=\, S_{e},
    \qquad x_{e}>0,
    \label{eq:star-heat-edge}
\end{align}
where the source, $S_{e}(t,x_{e})$, can have distinct values at the junction.
Energy balance supplies the vertex equation,
\begin{align}
    \mathrm{c}_{0}\,\partial_{t}T_{0}
    -\sum_{e=1}^{3}
        \left.K_{e}\,\partial_{x_{e}}T_{e}\right|_{x_{e}=0}
    \,=\,\mathrm{s}_{0}.
    \label{eq:star-heat-vertex}
\end{align}
The point capacity, $\mathrm{c}_{0}$, and source, $\mathrm{s}_{0}$, allow the junction itself to store and supply heat. Setting both to zero gives the usual Kirchhoff flux condition.
The semi-infinite edges isolate the central junction; we assume $T_{e}\to0$ as $x_{e}\to\infty$. Finite edges instead require conditions at their other ends, determined by the application.

Dedalus can already solve systems assembled in this form, including nonlinear Schr\"{o}dinger equations on a D-shaped graph \cite{burns_etal_2020}. However, specifying each edge equation and junction condition separately becomes cumbersome as the network and model grow. The same heat problem can be written as
\begin{align}
    \mathcal{C}\,\partial_{t}T
    -\div\!\left(\mathrm{K}\dat\grad T\right)
    \,=\,\mathcal{S},
    \qquad x\in G.
    \label{eq:abstract-heat}
\end{align}
Here $T$ is a continuous scalar field, the conductivity tensor $\mathrm{K}$ multiplies the gradient on edge $e$ by $K_{e}$, and the capacity and source include their point contributions:
\begin{align}
    \mathcal{C}=C+\mathrm{c}_{0}\,\delta_{0},
    \qquad
    \mathcal{S}=S+\mathrm{s}_{0}\,\delta_{0}.
\end{align}
On a finite graph, $\partial G$ denotes the designated boundary vertices. A homogeneous prescribed temperature condition takes the form $T|_{\partial G}=0$. The definitions below make~\eqref{eq:abstract-heat} reproduce both the edge equation and the vertex balance.

This example illustrates three advantages of abstraction. First, common operators allow definitions, analysis and numerical methods to be reused across different models and networks. Second, their defining identities connect local dynamics to global balances: the same divergence produces both the edge equations and the vertex flux conditions. Third, a covariant formulation makes additional modelling assumptions explicit. A preferred direction must enter through directional data, and an additional operator or constraint must specify its action on the fields. These requirements provide \emph{structural accountability}: extensions remain possible, but each must supply the mathematical structure needed to define it and account for its consequences.

\subsubsection{KdV, covariance and structural accountability}
\label{sec:kdv-covariance}

The linear KdV equation illustrates the need for explicit directional data. The edge equation
\begin{align}
    \partial_{t}u_{e}+\partial_{x_{e}}^{3}u_{e}=0
\end{align}
changes under reversal of the edge coordinate because
\begin{align}
    \partial_{x_{e}}\longrightarrow-\partial_{x_{e}}.
    \label{eq:kdv-coordinate-reversal}
\end{align}
A covariant formulation therefore includes a prescribed direction field. With the unit tangent, $\hat{e}$, pointing towards increasing $x_{e}$, define
\begin{align}
    \hat{E}|_{e}=\sigma_{e}\,\hat{e},
    \qquad
    \sigma_{e}\in\{-1,1\},
    \qquad
    (\hat{E}\dat\grad u)|_{e}
    =\sigma_{e}\,\partial_{x_{e}}u_{e}.
\end{align}
The components, $\sigma_{e}$, are constant along each edge. Equivalently, $\hat{E}=\grad X$ for an edgewise arclength field satisfying $\partial_{x_{e}}X_{e}=\sigma_{e}$; the values of $X$ need not agree at vertices. Reversing an edge coordinate changes the signs of both $\sigma_{e}$ and $\partial_{x_{e}}$, leaving $\hat{E}\dat\grad$ unchanged.
The formally skew-adjoint directional derivative,
\begin{align}
    \mathcal{D}_{\!\hat{E}\,}u
    \,=\,
    \frac12\left[
        \hat{E}\dat\grad u+\div(\hat{E}u)
    \right],
\end{align}
reduces to $\sigma_{e}\partial_{x_{e}}u_{e}$ in each edge interior. Its divergence term supplies the vertex contribution required by the skew identity.
The mixed system
\begin{align}
    \partial_{t}\mathcal{D}_{\!\hat{E}}u+\Delta w=0,
    \qquad
    w-\Delta u=0,
    \qquad
    \Delta=\div\grad,
    \label{eq:kdv-mixed-system}
\end{align}
defines a differentiated linear KdV model. The real fields, $u$ and $w$, are continuous at each vertex. Both equations hold as identities of functionals, with $w$ in the second equation interpreted as a regular edge density. Their edge equations are
\begin{align}
    \partial_{x_{e}}
    \left(\sigma_{e}\,\partial_{t}u_{e}
    +\partial_{x_{e}}w_{e}\right)=0,
    \qquad
    w_{e}=\partial_{x_{e}}^{2}u_{e}.
\end{align}
Thus $\sigma_{e}\partial_{t}u_{e}+\partial_{x_{e}}^{3}u_{e}$ is constant along each edge. Conditions that set this integration constant to zero recover the linear KdV equation with the prescribed direction.

The local Lagrangian density,
\begin{align}
    \mathcal{L}
    \,=\,
    \tfrac12\,\partial_{t}u\,\hat{E}\dat\grad u
    +\grad u\dat\grad w
    +\tfrac12 w^{2},
\end{align}
produces~\eqref{eq:kdv-mixed-system} when varied over the continuous fields. The construction makes the directional field, vertex terms and admissible variations explicit. Establishing well-posed evolution still requires an appropriate choice of endpoint and junction data. Structural accountability exposes those choices and their consequences within the formulation.

\section{Fields and differential operators on a metric graph}
\label{sec:network-fields}

Graph fields combine regular functions along edges with data supported at vertices. For scalar fields, these form the hierarchy
\begin{align}
    \mathsf{Continuous}
    \ \hookrightarrow\
    \mathsf{Piecewise}
    \ \hookrightarrow\
    \mathsf{Distributional}.
    \label{eq:field-regularity-hierarchy}
\end{align}
Continuous fields share a common value at each vertex. Piecewise fields allow independent incident-edge values, while distributional fields additionally contain Dirac measures and their derivatives. These distinctions specify what a field can represent. Additional conditions, such as a prescribed slope or a relation between values and derivatives, enter as equations.

Several constructions provide precedents. \citet{Friedman2004} define a graph calculus using edge and vertex measures, with divergence containing both edge derivatives and vertex fluxes. Their measure-valued construction does not supply a hierarchy of derivatives of vertex point masses. \citet{Kurasov1996} develops distribution theory for discontinuous test functions at an interface on the line. \citet[\S2.3]{Nortier2020} includes distributional derivatives in a construction for star and rose graphs. Distributional Hessian and div-div complexes on triangulations combine polynomial fields with Dirac measures \citep{Hu2025}. Variational formulations of general higher-order network operators describe a broad range of junction conditions \citep{Gregorio2020}. Here we organise regular edge fields, endpoint distributions and constraint operators for direct equation specification and spectral discretisation in Dedalus. Test fields define the operators and their vertex terms; Section~\ref{sec:network-discretization} describes how the resulting equations are enforced using polynomial expansions and tau corrections \cite{Burns2026a,Burns2026b}.

\subsection{Scalar and vector fields}
\label{sec:field-types}

A finite metric graph, $G=(V,E)$, comprises vertices connected by intervals. Each edge, $e\in E$, has length $\ell_{e}>0$ and coordinate $x_{e}\in[0,\ell_{e}]$. A designated set of vertices, $\partial G$, carries any external boundary conditions. The notation, $e\sim v$, counts incident edge ends; a self-loop contributes twice.

Independent of regularity, the symbols, $\mathsf S(G)$ and $\mathsf V(G)$, denote scalar and tangent-vector field types. The gradient and divergence exchange their tensor character:
\begin{align}
    \mathsf S(G)
    \xrightarrow{\ \grad\ }
    \mathsf V(G),
    \qquad
    \mathsf V(G)
    \xrightarrow{\ \div\ }
    \mathsf S(G).
    \label{eq:scalar-vector-mappings}
\end{align}
Their action on the regularity classes is specified below.

The unit tangent, $\hat{e}$, points towards increasing $x_{e}$. A tangent-vector field has the edge representation, $q=q_{e}\,\hat{e}$. Reversing the coordinate changes the signs of both the basis and its component, leaving the vector unchanged. Incident edges supply distinct tangent directions at a junction.

Combining regularity with tensor character gives the scalar field classes,
\begin{align}
    \mathsf{S}_{\mathrm{c}}(G)
    \ \hookrightarrow\
    \mathsf{S}_{\mathrm{pw}}(G)
    \ \hookrightarrow\
    \mathsf{S}_{\mathrm{d}}(G), \qquad \mathsf{V}_{\mathrm{pw}}(G)
    \ \hookrightarrow\
    \mathsf{V}_{\mathrm{d}}(G),
    \label{eq:scalar-vector-field-classes}
\end{align}
with $(\mathrm{c}, \mathrm{pw}, \mathrm{d})$ subscripts corresponding to continuous, piecewise and distributional respectively.
The class, $\mathsf{S}_{\mathrm{pw}}(G)$, contains regular edge fields with independent endpoint values. Regular fields are smooth along each closed edge; their numerical representations are finite polynomial expansions. The class, $\mathsf{S}_{\mathrm{c}}(G)$, additionally identifies all incident scalar values with a common value, $u(v)$. The distributional class, $\mathsf{S}_{\mathrm{d}}(G)$, adds the endpoint Dirac terms defined below. There is no default matching of vector values across a branching vertex. Comparing vectors in different incident tangent lines requires additional structure. Conservation of their summed flux is a separate, intrinsically defined condition.

Each regular edge component has a spectral representation,
\begin{align}
    f_{e}(x_{e})
    \,=\,
    \sum_{n=0}^{N_{e}-1}
        f_{en}\,P_{n}^{(e)}(x_{e}),
\end{align}
where $P_{n}^{(e)}$ is a degree-$n$ basis polynomial. Typically, all edges use the same polynomial family, scaled to their local interval lengths. A continuous-field declaration supplies the endpoint matching automatically. Further restrictions on fields are specified separately as constraints.

The outward derivative of an edge interval is
\begin{align}
    \partial_{n}
    \,=\,\widehat{B}_{ve}\,\partial_{x_{e}},
    \qquad
    \widehat{B}_{ve}
    \,=\,
    \begin{cases}
        -1, & x_{e}=0,\\
        +1, & x_{e}=\ell_{e}.
    \end{cases}
    \label{eq:outward-derivative}
\end{align}
For edges with distinct endpoints, these are the signs of the oriented incidence matrix, $B$. At a self-loop, the two endpoint signs remain separate in endpoint sums.

Graph integrals use arclength along each edge. The pairing is conjugate-linear in its first argument:
\begin{align}
    \big\langle\, f,g\,\big\rangle_{
    \!G}
    \,=\,\sum_{e\in E}\int_{e}\overline{f_{e}}\,g_{e}.
    \label{eq:graph-pairing}
\end{align}
The vector pairing uses the corresponding dot product. Distributional fields act on test fields that are smooth along each closed edge, with independent incident-edge traces. For equations tested against continuous scalar fields, terms acting on endpoint values combine at each vertex because all incident test values coincide.

An endpoint Dirac derivative is defined by
\begin{align}
    {\big\langle} \, \phi,\delta_{ve}^{(k)}\, \big\rangle_{\!G}
    \,=\,(-1)^{k}\,
        \partial_{n}^{k}\overline{\phi_{e}(v)},
    \qquad k=0,1,2,\ldots.
    \label{eq:endpoint-dirac-derivatives}
\end{align}
The case, $k=0$, is an endpoint measure. Higher derivatives act on the corresponding one-sided derivatives of the test field. For continuous scalar tests, all $\delta_{ve}^{(0)}$ at a vertex act as the same measure, $\delta_{v}$.

A distributional scalar field has the representation
\begin{align}
    F
    \,=\,f
    +\sum_{v\in V}\sum_{e\sim v}\sum_{k\ge 0}
        c_{ve}^{(k)}\,\delta_{ve}^{(k)}.
    \label{eq:distributional-field}
\end{align}
The regular part, $f=(f_{e})_{e\in E}\in\mathsf{S}_{\mathrm{pw}}(G)$, comprises the edge expansions defined above. Only finitely many singular coefficients, $c_{ve}^{(k)}$, are nonzero. Vector distributions retain the incident-edge tangent basis. A sparse dictionary stores the singular coefficients by vertex, edge end and derivative order. The endpoint labels are part of the distributional field and are retained during multiplication and differentiation. Contributions are combined only when the completed expression is paired with the chosen tests.

\subsection{Differentiation}
\label{sec:differentiation}

The edgewise gradient maps regular scalar fields to regular tangent-vector fields:
\begin{align}
    \grad\, : \, \mathsf{S}_{\mathrm{pw}}(G)
    & \ \longrightarrow \ \mathsf{V}_{\mathrm{pw}}(G),
    &
    (\grad u)|_{e}
    &\,=\,\hat{e}\,\partial_{x_{e}}u_{e}.
    \label{eq:network-gradient}
\end{align}
It acts in the same way on continuous scalars. For $q=q_{e}\hat{e}$,
\begin{align}
    q\dat\grad u=q_{e}\,\partial_{x_{e}}u_{e}.
\end{align}
For a piecewise scalar, this gradient differentiates the regular edge components; it does not itself add distributions for differences between endpoint values. 

The divergence is defined by its action on all piecewise scalar tests, $\phi\in\mathsf{S}_{\mathrm{pw}}(G)$, 
\begin{align} 
    \big \langle \, \phi,\div q\rangle_{G} 
    \,=\,-\langle\grad\phi,q\, \big\rangle_{\!G}. 
    \label{eq:network-divergence} 
\end{align} 
For a regular vector field, integration by parts gives 
\begin{align} 
    \div q 
    \, = \, 
    \left(\partial_{x_{e}}q_{e}\right)_{e\in E} 
    -\sum_{v\in V}\sum_{e\sim v} 
        \widehat{B}_{ve}\,q_{e} (v)\,\delta_{ve}^{(0)}. 
    \label{eq:network-divergence-endpoints} 
\end{align} 
Integrated against continuous scalar test functions $\delta_{ve}^{(0)} \to \delta_{v}$. 
Thus 
\begin{align}
\div \, : \,  \mathsf{V}_{\mathrm{pw}}(G) \ \longrightarrow \  \mathsf{S}_{\mathrm{d}}(G). 
\end{align}

For a continuous scalar $u$, the distributional Laplacian $\Delta=\div\grad$ has the following representation when evaluated on continuous scalar tests:
\begin{align}
    \Delta u
    \,=\,
    \left(\partial_{x_{e}}^{2}u_{e}\right)_{e\in E}
    -\sum_{v\in V} \delta_{v}\, \sum_{e\sim v}\partial_{n}u_{e}(v).
    \label{eq:graph-laplacian}
\end{align}

Piecewise regular coefficients multiply endpoint distributions through the product rule:
\begin{align}
    f\,\delta_{ve}^{(k)}
    \,=\,
    \sum_{j=0}^{k}
        (-1)^{j}\binom{k}{j}\,
        \partial_{n}^{j} f_{e}(v)\,
        \delta_{ve}^{(k-j)}.
    \label{eq:distribution-product}
\end{align}

\subsubsection{Expressions and equations}

There is an important subtlety in evaluating Dirac distributions at a vertex. A piecewise test function can have different endpoint values on its incident edges, whereas a continuous test function
has one common value. For example, at a vertex, $v$, joining edges $1$ and $2$,
\begin{align}
    \big\langle\,
        \phi,\, a_{1}\delta_{v1}^{(0)}+a_{2}\delta_{v2}^{(0)}\,
    \big\rangle_{G}
    \,=\,
    a_{1}\,\overline{\phi_{1}(v)}
    +a_{2}\,\overline{\phi_{2}(v)} 
\end{align}
If $\phi$ is continuous at $v$, the result becomes 
\begin{align}
    \big\langle\,
        \phi,\, a_{1}\,\delta_{v1}^{(0)}+a_{2}\,\delta_{v2}^{(0)}\,
    \big\rangle_{G}
    \,=\,
    (a_{1} + a_{2}) \,\overline{\phi(v)}
\end{align}
The two endpoint contributions then act as a single vertex measure, and cancel if $a_{1}+a_{2}=0$. Multiplication by a discontinuous coefficient can weight these contributions differently and remove that cancellation. We therefore retain the individual endpoint distributions throughout the algebra and combine their contributions when the complete equation is evaluated on continuous tests.

With the multiplication defined above, the divergence satisfies the Leibniz rule,
\begin{align}
    \div(Kq)
    \,&=\,K\,\div q + q \dat \grad K,
    \label{eq:graph-leibniz}
\end{align}
for $K\in\mathsf S_{\mathrm{pw}}(G)$ and
$q\in\mathsf V_{\mathrm{pw}}(G)$.
This identity holds before combining endpoint contributions. For discontinuous coefficients, multiplication must precede combination of endpoint contributions. For a continuous test field, $\phi$,
\begin{align}
    \big\langle\,
        \phi,\,
        K\,\sum_{e\sim v}c_{ve}^{(0)} \, \delta_{ve}^{(0)}
    \,\big\rangle_{G}
    \,=\,
    \overline{\phi(v)}\, 
    \sum_{e\sim v}K_{e}(v)\, c_{ve}^{(0)}.
\end{align}
Dedalus performs the final combination assuming continuous test fields when assembling the complete equations into the system of spectral coefficients and tau corrections. Continuity of test values does not require continuity of material coefficients or identify incident-edge derivative traces. We leave this final assembly convention implicit in our scalar equations, using $\delta_{ve}^{(0)}$ for individual endpoint contributions and $\delta_{v}$ for the action on continuous tests.

\subsection{Operator composition}
\label{sec:operator-composition}

We demonstrate how higher-order derivatives work via a case study of the beam equation.
Each edge has a one-dimensional tangent space. An intrinsic tensor of order $r$ has the form
\begin{align}
    T|_{e}=T_{e}\,\hat{e}^{\otimes r}.
\end{align}
Contractions remove pairs of tangent factors because
$\tr(\hat{e}\otimes\hat{e})=\hat{e}\dat\hat{e}=1$.
Even-order tensors therefore reduce to scalars and odd-order tensors to vectors, without losing any edge components. More specifically, all intrinsic rank-2 tensors act by scalar multiplication, 
\begin{align}
    \mathrm{K}|_{e}\, = \, K_{e}\,\hat{e}\otimes\hat{e}, \qquad \mathrm{K}\dat q \, = \, K\,q. 
\end{align} 
For a piecewise regular tangent-vector field $q=q_{e}\hat{e}$, the edgewise gradient is 
\begin{align} 
    (\grad q)|_{e} 
    \,=\, (\partial_{x_{e}}q_{e})\,\hat{e}\otimes\hat{e}. 
    \label{eq:edge-vector-gradient} 
\end{align} 
The scalar representation of the edgewise  Hessian is 
\begin{align} 
    \mathcal H u 
    \,=\,\tr(\grad\grad u),
    \qquad
    (\mathcal H u)_{e}=\partial_{x_{e}}^{2}u_{e}.
    \label{eq:edge-hessian}
\end{align}
The operator, $\mathcal H$, contains only regular curvature; $\Delta=\div\grad$ contains additional vertex terms in~\eqref{eq:graph-laplacian}.

A displacement, $u\in\mathsf S_{\mathrm c}(G)$, has the beam Lagrangian density
\begin{align}
    \mathcal{L}
    \, = \, 
    \frac{\rho}{2}\,|\partial_{t}u|^{2} 
    -\frac{K}{2}\,|\mathcal H u|^{2} + f\,u, 
    \label{eq:beam-lagrangian} 
\end{align} 
Here $f$ is the applied force density, while $\rho,K\in\mathsf S_{\mathrm{pw}}(G)$ are the linear density and bending stiffness, respectively. Variation of the  action gives
\begin{align}
    \rho\,\partial_{t}^{2}u+\mathcal H^{*}(K\mathcal H u)=f.
    \label{eq:beam-abstract}
\end{align}
Within an edge, 
\begin{align}
    \rho_{e}\,\partial_{t}^{2}u_{e}
    +\partial_{x_{e}}^{2} \left(K_{e}\,\partial_{x_{e}}^{2}u_{e}\right) 
    \,=\,f_{e}. 
    \label{eq:beam-edge} 
\end{align} 
The adjoint is defined on piecewise scalar tests by
\begin{align} 
    \big \langle\, \phi,\mathcal H^{*}m\, \big \rangle_{\!G} 
    \,=\,\big \langle\, \mathcal H\phi,m\, \big \rangle_{\!G} \qquad \text{where} \qquad m \, = \, K\mathcal H u
    \label{eq:hessian-adjoint-definition} 
\end{align}
is the bending moment. Two integrations by parts give
\begin{align}
    \mathcal H^{*}m
    \,=\,
    \left(\partial_{x_{e}}^{2}m_{e}\right)_{e\in E}
    -\sum_{v\in V}\sum_{e\sim v}
        \left[
            \partial_{n}m_{e}(v)\,\delta_{ve}^{(0)}
            +m_{e}(v)\,\delta_{ve}^{(1)}
        \right].
    \label{eq:beam-adjoint}
\end{align}
The edge expression is fixed. Relations between admissible endpoint variations determine which singular contributions remain independent. For a continuous field with no further constraints, the common vertex value of $\phi$ and its independent incident-edge derivatives give, in the absence of vertex forcing,
\begin{align}
    \sum_{e\sim v}\partial_{n}m_{e}(v)=0,
    \qquad
    m_{e}(v)=0
    \quad\text{for every }e\sim v.
    \label{eq:beam-natural-junction}
\end{align}
Other junction conditions can be imposed by adding explicit constraints and their unknown reaction sources.

The endpoint trace operators provide the missing piece, 
\begin{align}
    (\mathcal T_{k}u)_{ve}
    \,=\,\partial_{n}^{k}u_{e}(v),
    \qquad k=0,1,2,\ldots,
    \label{eq:endpoint-traces}
\end{align}
with adjoints defined using the standard Hermitian pairing on endpoint arrays,
\begin{align}
    \mathcal T_{k}^{*}\lambda
    \,=\,(-1)^{k}
        \sum_{v\in V}\sum_{e\sim v}
        \lambda_{ve}\,\delta_{ve}^{(k)}.
    \label{eq:trace-adjoints} 
\end{align} 
A linear constraint is specified by 
\begin{align} 
    \mathcal{C} u=g, 
    \qquad \mathcal{C}=\sum_{k=0}^{r}A_{k}\mathcal T_{k}, 
    \qquad \mathcal{C}^{*}=\sum_{k=0}^{r}\mathcal T_{k}^{*}A_{k}^{*}. 
    \label{eq:constraint-operator} 
\end{align} 
The matrices, $A_{k}$, select and combine endpoint data. They can impose a rule throughout the graph or on a designated set of vertices. The constraint rows are chosen independently on the declared field class; continuity already supplied by $\mathsf S_{\mathrm c}(G)$ is not imposed again. 

The constrained beam system is 
\begin{align} 
    \rho\,\partial_{t}^{2}u 
    +\mathcal H^{*}(K\mathcal H u) + \mathcal{C}^{*}\lambda 
    &=f, 
    & 
    \mathcal{C} u&=g. 
    \label{eq:constrained-beam} 
\end{align} 
The purpose of this example is to show how explicit constraints supplement the conditions supplied by an adjoint. With the field class fixed, the adjoint gives natural endpoint and junction balances, just as divergence gives Kirchhoff's law in the heat equation. To impose a different condition, e.g. \cite{Olver2022}, $\mathcal{C}u=g$ supplies the constraint and $\mathcal{C}^{*}\lambda$ supplies an unknown reaction, with ``slack variable'' $\lambda$, in the field equation. This is very analogous to the tau method approach used in Dedalus to enforce strong-form systems of equations. The endpoint equations then determine the reaction in place of the corresponding natural condition, while the remaining balances still hold. For example, $\mathcal{C}=\mathcal T_{1}$ and $g=0$ prescribe zero slopes while allowing nonzero endpoint moments. The same ideas allow more general options in the KdV example as well.

Table~\ref{tab:beam-vertex-implementations} gives representative endpoint and junction conditions obtained by choosing $\mathcal{C}$ in~\eqref{eq:constrained-beam} \citep{Gregorio2019,Gregorio2020}. All use the same edge operator and distributional adjoint.

\begin{table}
    \centering
    \small
    \renewcommand{\arraystretch}{1.35}
    \begin{tabular}{@{}p{0.56\linewidth}p{0.38\linewidth}@{}}
        \toprule
        Homogeneous endpoint or vertex conditions
        & Constraint expression, $\mathcal{C} u$ \\
        \midrule
        Prescribed endpoint value:\newline
        $u=0$, $m=0$
        & $(\mathcal T_{0}u)_{ve}$ \\
        Prescribed endpoint slope:\newline
        $\partial_{n}u=0$, $\partial_{n}m=0$
        & $(\mathcal T_{1}u)_{ve}$ \\
        Mixed endpoint condition, with real $(a,b)\ne(0,0)$:\newline
        $a u+b\,\partial_{n}u=0$,\quad
        $a m+b\,\partial_{n}m=0$
        & $a\,(\mathcal T_{0}u)_{ve}+b\,(\mathcal T_{1}u)_{ve}$ \\
        Coupled junction:\newline
        $\sum_{e\sim v}\partial_{n}u_{e}=0$; all $m_{e}$ equal;\newline
        $\sum_{e\sim v}\partial_{n}m_{e}=0$
        & $\displaystyle\sum_{e\sim v}(\mathcal T_{1}u)_{ve}$ \\
        \bottomrule
    \end{tabular}
    \caption{Examples obtained by setting the indicated constraint
    expression to zero and including its adjoint reaction in
    \eqref{eq:constrained-beam}. The displacement is continuous and
    $f$ has no vertex-supported terms. The first three rows apply at
    degree-one endpoints; the last applies at a junction.}
    \label{tab:beam-vertex-implementations}
\end{table}

\section{Computational discretization}
\label{sec:network-discretization}

The graph calculus forms an interface layer over the existing Dedalus infrastructure \citep{burns_etal_2020}. Users declare graph fields and enter equations in the abstract forms illustrated by the heat, KdV and beam examples. The interface translates these expressions and their trace constraints into edge and vertex equations, then assembles the polynomial coefficients, vertex unknowns and tau corrections.
These models use the standard Dedalus problem classes: initial-value problems, linear and nonlinear boundary-value problems, and linear eigenvalue problems. Dedalus supplies the time integrators, Newton iterations and eigenvalue algorithms, while the graph interface supplies the fields, operators and vertex coupling.
For linear eigenmodes of a time-independent system, the ansatz, $u \propto e^{\sigma t}$, replaces $\partial_{t} \to \sigma$. Reducing higher-order time derivatives to first order where needed gives Dedalus's standard generalized eigenvalue form.

The code boxes below illustrate the interface for the static beam problem in Section~\ref{sec:operator-composition}, using supplied network data.
Three unit-length beams have unit bending stiffness, unequal distributed loads, and clamped outer ends.
The central displacement is continuous and the outward slopes sum to zero, using the coupled-junction condition in Table~\ref{tab:beam-vertex-implementations}.
All quantities are nondimensional. Each box introduces the code for the corresponding concept.

\subsection{Network bases and domains}
\label{sec:cardinal-network-domains}

The cardinal basis allows discrete labels to participate in the same field representation as continuous coordinates.
For the finite edge set $E$, define
\begin{equation}
    \chi_i(e)=\delta_{ie},
    \qquad i,e\in E.
    \label{eq:edge-cardinal-basis}
\end{equation}
Let $s\in[0,1]$ denote a reference coordinate on each edge, and take a common polynomial interval basis $\{\phi_n\}_{n=0}^{N-1}$.
The product expansion
\begin{equation}
    u(e,s)=\sum_{i\in E}\sum_{n=0}^{N-1}
        u_{in}\,\chi_i(e)\,\phi_n(s)
    \label{eq:cardinal-network-expansion}
\end{equation}
represents the edge restrictions of a physical field together as a single Dedalus field, indexed by the edge label and the interval coordinate.
Each edge retains its own polynomial coefficients $u_{in}$.
The physical coordinate on edge $e$ is $x_e=\ell_e s$, so
\begin{equation}
    \partial_{x_e}=\ell_e^{-1}\partial_s,
    \qquad
    \mathrm d x_e=\ell_e\,\mathrm d s.
    \label{eq:edge-physical-scaling}
\end{equation}
These length factors also enter the evaluation of endpoint derivatives.

The representation on the finite set is exact: a function on $E$ has its values as cardinal coefficients, so the transform along this direction is the identity.
The polynomial truncation acts along the interval coordinate.
The identity $\chi_i\chi_j=\delta_{ij}\chi_i$ makes multiplication by an edge-dependent scalar diagonal in the cardinal direction, while differentiation and polynomial transforms act along each interval.
Edge-dependent lengths and material properties therefore fit directly into the usual coefficient operations.

Analogous cardinal spaces index vertices and incident endpoints.
Sparse maps between these spaces encode incidence, restriction and sums over incident edges.
Together with endpoint evaluation of the interval expansions, these maps connect edge fields to vertex data.
The declared field regularity and the supplied vertex equations specify continuity and other coupling conditions.
The \texttt{NetworkBasis} interface packages the cardinal and polynomial bases for use with ordinary Dedalus fields and problem classes.

This construction applies to any finite index set; networks supply a particular collection of spaces and maps.
It brings discrete coupling into Dedalus's existing symbolic equations, spectral operators and distributed field storage.

The supplied \texttt{vertex\_ids} form an ordered table of vertex labels; \texttt{endpoints} lists ordered pairs of indices into that table, and \texttt{lengths} gives the corresponding edge lengths.
The mapping \texttt{vertex\_groups} assigns names to sets of vertex-table indices, using \texttt{clamped} for the supports and \texttt{coupled} for the junctions in the beam example.
Group names alone impose no conditions.
The distributor manages field storage, and the network basis combines the edge cardinal basis with $N$ retained polynomial modes per edge.
In the code examples, \texttt{d3} denotes the Dedalus public interface.

\noindent\begin{minipage}{\linewidth}
\begin{lstlisting}[style=networkpython,label={lst:beam-network}]
network = d3.Network(vertex_ids=vertex_ids, endpoints=endpoints, lengths=lengths, vertex_groups=vertex_groups)
dist = d3.Distributor(network, dtype=np.float64)
basis = d3.NetworkBasis(network, size=N)
\end{lstlisting}
\end{minipage}

\subsection{Cohomology and spectral discretisation}
\label{sec:compatible-spectral}

The polynomial representation on a graph satisfies the underlying cohomology exactly. That is, take an edgewise polynomial in $\mathsf{V}_{\mathrm{pw}}(G)$,
\begin{align} 
    q|_{e} \, = \, \sum_{n=0}^{N-1} q^{(e)}_{n} \, p^{(e)}_{n}(x) \, \hat{e},
\end{align}
where $p^{(e)}_{n}(x)$ is the degree-$n$ polynomial basis element on $e \in E$.
There is no nontrivial intrinsic local $\operatorname{curl}$ operator on a graph. However, $q$ still has a Hodge decomposition of the form,
\begin{align} 
    q \, = \, \grad f + h,
\end{align}
where $f \in \mathsf{S}_{\mathrm{c}}(G)$,  
\begin{align}
    f|_{e} \, = \, \sum_{n=0}^{N} f^{(e)}_{n}\, p^{(e)}_{n}(x),
\end{align}
which is unique up to a global gauge constant $f \to f + f_{0}$.
Both expansions use the same local polynomial basis without loss of generality. The remaining field, $h$, is constant on each edge and satisfies the Kirchhoff condition at every vertex, \textit{i.e.},
\begin{align}
    \partial_{x_{e}} h_{e}=0,
    \qquad
    \sum_{e\sim v}\widehat{B}_{ve}\,h_{e}(v)=0.
\end{align}
The field, $h$, is orthogonal to the gradient of every continuous scalar, 
\begin{align}
    \big \langle \, \grad \phi \,,  \, h \, \big \rangle_{\!G} \, = \, 0,
\end{align}
for all $\phi \in \mathsf{S}_{\mathrm{c}}(G)$, and carries all the circulation of $q$. That is, for any closed loop, $\gamma \in \pi_{1}(G)$, 
\begin{align}
\Gamma(\gamma) \, = \, \int_{\gamma} q \dat \mathrm{d} \ell \, = \, \int_{\gamma} h \dat \mathrm{d} \ell.
\end{align}
A graph with a single connected component contains $|\!\operatorname{cycles}(G)| = |E|-|V|+1$ linearly independent cycles, 
\begin{align}
    h \, = \, \sum_{\gamma \in \operatorname{cycles}(G)} \Gamma(\gamma)
    \, h(\gamma) \qquad \text{where} \qquad \int_{\gamma} h(\gamma')\dat \mathrm{d} x \, = \, \delta_{\gamma,\gamma'},
\end{align}
where $\operatorname{cycles}(G)$ is a well-defined linearly independent basis of cycles. 

Expressing vectors and scalars in different polynomial bases changes their coefficients but leaves the cohomology unchanged. The point here is that finite element and spectral implementations share the same compatible polynomial spaces. 
Dedalus enforces the resulting strong-form edge equations with explicit tau corrections in spectral coefficient space, combined with various vertex equations.

\subsection{Fields}
\label{sec:network-field-interface}

An ordinary Dedalus field on the network basis stores the displacement on all edges.
Declaring \texttt{regularity=`continuous'} identifies its incident endpoint values at each vertex; it leaves the slopes independent.
The bending stiffness, called \texttt{EI} in the script, is constant.
The load uses only \texttt{basis.edges}, so it has one value per edge and is constant along each interval.
The local cardinal grid supplies the edge indices $0$, $1$, and $2$, giving loads $24$, $48$, and $72$.
The \texttt{[`g']} accessor sets a field's grid values.
We also retain the named vertex groups for the constraints below.

\noindent\begin{minipage}{\linewidth}
\begin{lstlisting}[style=networkpython,label={lst:beam-fields}]
# Displacement and coefficients
u = dist.Field(name='u', bases=basis, regularity='continuous')
u['g'] = 0

clamped = network.vertex_groups['clamped']
selected_junctions = network.vertex_groups['coupled']
EI = 1.
f = dist.Field(name='f', bases=basis.edges)
edge = dist.local_grid(basis.edges)
f['g'] = 24 * (edge + 1)
\end{lstlisting}
\end{minipage}

\subsection{Equations and vertex constraints}
\label{sec:network-problem-specification}

We specify the static beam balance as a linear boundary-value problem.
The equation namespace supplies the fields and coefficients, while \texttt{H} and \texttt{H.T} are available directly in equation strings.
They denote the edge curvature and its distributional transpose, including the endpoint moments and forces in~\eqref{eq:beam-adjoint}.

\noindent\begin{minipage}{\linewidth}
\begin{lstlisting}[style=networkpython,label={lst:beam-equation}]
# Problem
problem = d3.LBVP([u], namespace=locals())
problem.add_equation('H.T(EI*H(u)) = f')
\end{lstlisting}
\end{minipage}

The support and joint kinematics are supplied through \texttt{endpoint}, the trace operator in~\eqref{eq:endpoint-traces}.
Order \texttt{deriv=0} gives endpoint values and \texttt{deriv=1} gives outward physical slopes.
The argument \texttt{on} selects vertices; \texttt{sum=True} sums incident endpoint contributions separately at each selected vertex.
Thus the first two calls below clamp the outer ends, while the third imposes the chosen junction slope sum.
Each \texttt{add\_vertex\_constraint} call both imposes $\mathcal C u=g$ and introduces its unknown reaction $\mathcal C^*\lambda$ into the beam balance, as in~\eqref{eq:constrained-beam}.

\noindent\begin{minipage}{\linewidth}
\begin{lstlisting}[style=networkpython,label={lst:beam-constraints}]
# Vertex constraints add their unknown adjoint reactions to the beam balance.
problem.add_vertex_constraint('endpoint(u, deriv=0, on=clamped) = 0')
problem.add_vertex_constraint('endpoint(u, deriv=1, on=clamped) = 0')
problem.add_vertex_constraint('endpoint(u, deriv=1, on=selected_junctions, sum=True) = 0')
\end{lstlisting}
\end{minipage}

The equivalent call is \texttt{problem.add\_equation(..., vertex\_constraint=True)}.
Edge tau corrections are generated automatically and remain accessible through \texttt{problem.tau\_fields}; custom explicit formulations can disable this generation with \texttt{tau=False} in \texttt{add\_equation}.

\subsection{The assembled system}
\label{sec:assembled-system}

Dedalus discretizes the edge equations using its spectral tau method \cite{burns_etal_2020}.
Unknown tau coefficients mutiply prescribed polynomial corrections and are solved for alongside the field coefficients, allowing junctions to be imposed on the truncated expansions.

For a linear solve, we let $\mathbf z_e$ collect the expansion and tau coefficients of all fields on edge $e$.
$\mathbf U$ collects the additional unknowns associated with vertex conditions and global constraints, including shared vertex values and constraint reactions.
Writing $\mathbf z=(\mathbf z_{e})_{e\in E}$, the assembled equations have the block form
\begin{equation}
    \begin{pmatrix}
        \mathbf A & \mathbf B\\
        \mathbf C & \mathbf D
    \end{pmatrix}
    \begin{pmatrix}
        \mathbf z\\
        \mathbf U
    \end{pmatrix}
    =
    \begin{pmatrix}
        \mathbf f\\
        \mathbf g
    \end{pmatrix},
    \qquad
    \mathbf A=\operatorname{diag}_{e\in E}(\mathbf A_{e}).
    \label{eq:assembled-block-system}
\end{equation}
The first block row contains the edge equations and endpoint continuity conditions.
Each $\mathbf A_{e}$ acts on the coefficients belonging to one edge, while $\mathbf B$ connects its endpoint values to the shared vertex unknowns.
The second block row imposes vertex balances, prescribed constraints and any global conditions.
The block $\mathbf C$ contains contributions from the edge unknowns and $\mathbf D$ contains the coefficients from the vertex, reaction and global unknowns in $\mathbf U$.
The right-hand sides contains the known terms for the given linear solve.
The block-diagonal structure of $\mathbf A$ follows from coupling distinct edges only through these retained unknowns.

For the beam example, the assembled system includes the fourth-order edge equations and the unknown support and junction reactions.
The usual Dedalus solver interface builds and solves this system:

\noindent\begin{minipage}{\linewidth}
\begin{lstlisting}[style=networkpython,label={lst:beam-solver}]
# Solver
solver = problem.build_solver()
solver.solve()
\end{lstlisting}
\end{minipage}

The chosen loads admit a polynomial exact solution, providing a check of the assembled beam problem.
Here \texttt{x} is the physical coordinate measured from the central junction along each unit edge.
The final box reports the maximum displacement error on the grid for this three-edge example.

\noindent\begin{minipage}{\linewidth}
\begin{lstlisting}[style=networkpython,label={lst:beam-check}]
# Check against the exact solution
x = basis.local_physical_grid(dist)
exact = 2*(1-x*x)**2 + (edge-1)*(x**4 - 1.5*x**3 + 0.5*x)
error = np.max(np.abs(u['g'] - exact))
print(f'Maximum beam displacement error: {error:.6e}')
\end{lstlisting}
\end{minipage}

\section{Parallelization over subgraphs}
\label{sec:solution-algorithms}

We distribute the network over subgraphs and reduce the assembled equations to a coupled system on their interfaces.
After solving this system, the eliminated unknowns are recovered locally within each subgraph.

\subsection{Network partitioning}
\label{sec:subgraph-partitioning}

\subsubsection{Subgraphs and interface unknowns}

We partition the edge set into disjoint subsets,
\begin{equation}
    E=\bigcup_{p=1}^{P}E_p,
    \label{eq:edge-partition}
\end{equation}
and let $V_p$ contain the endpoints of the edges in $E_p$.
The subgraph $G_p=(V_p,E_p)$ is assigned to one parallel process, which holds the coefficients and equations associated with its edges.
Each edge belongs to exactly one subgraph, while vertices may be shared by several subgraphs.
A subgraph may contain many edges and does not need to be connected.

A vertex in $V_p$ is internal to subgraph $p$ if all its incident edges belong to $E_p$; denote the set of these vertices by $I_p$.
Vertices incident to edges in more than one subgraph form the interface $\Gamma$, with $\Gamma_p=V_p\cap\Gamma$ the interface of subgraph $p$.

If each junction carries several unknowns--for example, the concentrations of two reacting species--the interface retains whichever of those unknowns couple adjacent subgraphs. 
Those are determined by the vertex conditions, so they are not automatically common endpoint values.
A global constraint can also couple otherwise separate subgraphs. 
For example, prescribing a network-wide mean may introduce a single multiplier that affects equations throughout the network. 
That multiplier stays in the reduced system, and each subgraph contributes to the global constraint.

\subsubsection{Work estimates}

We estimate local work by assigning a weight to each edge.
Equal weights balance edge counts and are appropriate when the equations and resolutions are comparable across edges.
For heterogeneous discretizations, the number of expansion and tau coefficients provides an inexpensive initial estimate of edge work.
For the two-field, second-order problems considered below, an edge with $N_{e}$ expansion coefficients per field has weight $2(N_{e}+2)$, including its tau coefficients.
Length weights are useful when the resolution is approximately proportional to edge length, as in the heat-conduction examples below.

We also assign weights to junctions, estimating the number of scalar interface unknowns associated with each junction.
For the same two-field problems, each junction has weight two.

These estimates do not account fully for the cost of factoring the coupled subgraph systems, which also depends on their connectivity and equation structure.
A smaller interface can increase local work imbalance or factorization cost, so interface size alone is insufficient to assess a partition.
We therefore consider the complete solver cost, including setup and repeated solves, when choosing a partition for an application.
Whole edges remain indivisible, which can limit the achievable balance when a few edges dominate the work.

\subsubsection{Partition construction}

For general networks, we use Mt-KaHyPar (Multi-Threaded Karlsruhe Hypergraph Partitioner)~\citep{Gottesbren2024a} to construct a $P$-subgraph partition.
We construct a hypergraph in which the network edges become vertices and the network junctions become hyperedges.
For example, three edges meeting at a junction become three vertices joined by one hyperedge. Mt-KaHyPar divides these vertices into $P$ groups, balancing their total weight while penalizing hyperedges that cross between groups.
In the original network, this balances the work across processes while reducing the number of processes sharing each junction.
We use the weighted connectivity objective: a junction contained within one subgraph contributes nothing, while a junction shared by two subgraphs contributes its weight, one shared by three contributes twice its weight, and so on.

If a spatial embedding of the network is provided, we also offer spatial partitioning, which divides the network geometrically while balancing edge weights.
Other available constructions grow subgraphs through adjacent edges or assign edges solely to balance work.
Users can instead supply their own edge assignments.
Once selected, the assignment remains fixed throughout the solve.

\subsection{Subgraph interior elimination}
\label{sec:subgraph-interior-elimination}

We collect the edge expansion, tau coefficients and the internal vertex and reaction unknowns of subgraph $p$ into $\mathbf y_p$.
Let $\boldsymbol\lambda$ contain the retained interface unknowns and any global unknowns.
Reordering the assembled equations gives
\begin{equation}
    \begin{pmatrix}
        \mathbf H & \mathbf F\\
        \mathbf G & \mathbf J
    \end{pmatrix}
    \begin{pmatrix}
        \mathbf y\\
        \boldsymbol\lambda
    \end{pmatrix}
    =
    \begin{pmatrix}
        \mathbf b\\
        \mathbf c
    \end{pmatrix},
    \qquad
    \mathbf H=\operatorname{diag}_{p=1}^{P}(\mathbf H_p),
    \label{eq:subgraph-block-system}
\end{equation}
where $\mathbf y=(\mathbf y_p)_{p=1}^{P}$.
Each block $\mathbf H_p$ contains the interior equations and unknowns of one subgraph.
The remaining blocks couple these interiors to the retained unknowns and impose the interface conditions.

Assuming each $\mathbf H_p$ is invertible, eliminating the interior unknowns yields
\begin{equation}
    \mathbf S_\Gamma\boldsymbol\lambda=\mathbf r_\Gamma,
    \qquad
    \mathbf S_\Gamma=\mathbf J-\mathbf G\mathbf H^{-1}\mathbf F,
    \qquad
    \mathbf r_\Gamma=\mathbf c-\mathbf G\mathbf H^{-1}\mathbf b.
    \label{eq:subgraph-schur-system}
\end{equation}
We factor each subgraph matrix $\mathbf H_p$ as a single sparse system containing the edge coefficients, tau variables and internal vertex and reaction unknowns.
These factorizations are computed independently across processes and reused to construct the interface system and reconstruct the local solution.
Only the retained unknowns enter the coupled interface solve.
The inverse notation denotes applications of local factorizations.

\subsection{Interface solution and local reconstruction}
\label{sec:interface-solution-reconstruction}

We assemble the interface system $\mathbf S_\Gamma\boldsymbol\lambda=\mathbf r_\Gamma$ by combining the subgraph contributions to the interface conditions and any global constraints.
Each interface equation is assigned to one process, which collects the contributions to that matrix row.

We solve this system using restarted GMRES with diagonal right preconditioning.
Each process stores its assigned matrix rows and the corresponding entries of the Krylov vectors.
To evaluate a matrix--vector product, it obtains the additional vector entries required by those rows from the processes holding them.
Inner products and residual norms are computed by summing contributions across processes.
We form the preconditioner from the assembled interface diagonal.
We accept an iterative solution when the explicitly evaluated residual satisfies
\begin{equation}
    \lVert\mathbf r_\Gamma-\mathbf S_\Gamma\boldsymbol\lambda\rVert_2
    \leq
    \max\!\left(\varepsilon_{\mathrm{abs}},
    \varepsilon_{\mathrm{rel}}\lVert\mathbf r_\Gamma\rVert_2\right),
    \label{eq:interface-residual-tolerance}
\end{equation}
where $\varepsilon_{\mathrm{abs}}$ and $\varepsilon_{\mathrm{rel}}$ are the absolute and relative solver tolerances.
The interior unknowns are then recovered independently through
\begin{equation}
    \mathbf y_p=\mathbf H_p^{-1}
    \bigl(\mathbf b_p-\mathbf F_p\boldsymbol\lambda\bigr),
    \qquad p=1,\ldots,P,
    \label{eq:subgraph-reconstruction}
\end{equation}
where $\mathbf b_p$ and $\mathbf F_p$ are the block rows associated with subgraph $p$.
Only retained unknowns coupled to that subgraph contribute to $\mathbf F_p\boldsymbol\lambda$.
Reconstruction recovers both the edge fields and the internal vertex and reaction unknowns, using the local factorizations already computed during elimination.

This reduction and reconstruction are algebraically equivalent to solving the full assembled system whenever the required inverses exist.
They introduce no additional discretization approximation; numerical differences arise from roundoff and the tolerances used in the linear solves.

\subsection{Application to different problem classes}
\label{sec:reduction-problem-classes}

We use Dedalus's time integration, Newton and eigenvalue algorithms~\citep{burns_etal_2020}, applying the subgraph reduction to their linear solves.
The reduction acts on the assembled system for steady linear problems, the Jacobian for Newton corrections, and the shifted systems used in shift-invert eigenvalue calculations.
Newton corrections include the linearized vertex conditions.

For initial value problems, we reduce the implicit system at each timestep or stage.
Edge and vertex time-derivative terms enter the system together, while algebraic vertex conditions remain constraints.

We reuse local factorizations and the reduced interface operator while the system matrix remains unchanged.
When several Runge--Kutta stages have the same implicit matrix, they also share these factorizations and the interface operator.
This avoids storing and constructing separate copies for each stage.
Changes to the timestep, implicit coefficients, Newton iterate or eigenvalue shift may require rebuilding them.

\section{Numerical verification and parallel performance}
\label{sec:numerical-examples}

We now assess the numerical accuracy and parallel performance of the network solver.
We measure convergence against analytical and refined numerical reference solutions, and compare distributed solutions with direct solutions of the assembled system.
Residuals of the edge equations and vertex conditions provide an additional check that the computed fields satisfy the prescribed problem.
We then measure strong and weak scaling on networks with contrasting connectivity.

\subsection{Heat conduction in a network of rods}
\label{sec:rod-heat-conduction}

We first consider steady heat conduction through a network of rods that exchange heat with their surroundings.
The field $u_{e}$ denotes temperature relative to a uniform ambient temperature.
Each rod has constant coefficients $K_{e}>0$ and $A_{e}>0$, describing axial heat conduction and linear heat loss, respectively, and satisfies
\begin{equation}
    -K_{e}\frac{\mathrm d^{2} u_{e}}{\mathrm d x_{e}^{2}}+A_{e}u_{e}=0,
    \qquad 0<x_{e}<\ell_{e}.
    \label{eq:rod-heat-equation}
\end{equation}
We impose continuity of temperature at every vertex and conservation of heat flux at vertices where the temperature is not prescribed:
\begin{gather}
    u_{e}(v)=u(v), \qquad e\sim v,
    \label{eq:rod-junction-conditions1}
    \\
    \sum_{e\sim v}K_{e} \,\partial_nu_{e}(v)=0.
    \label{eq:rod-junction-conditions2}
\end{gather}

For an edge $e$ with endpoints $v$ and $w$ (corresponding to $x_{e}=0$ and $x_{e}=\ell_{e}$, respectively), we write $\kappa_{e}=\sqrt{A_{e}/K_{e}}$ and the exact temperature profile is
\begin{equation}
    u_{e}(x_{e})
    =u(v)\frac{\sinh\!\bigl(\kappa_{e}(\ell_{e}-x_{e})\bigr)}
               {\sinh(\kappa_{e}\ell_{e})}
    +u(w)\frac{\sinh(\kappa_{e}x_{e})}
               {\sinh(\kappa_{e}\ell_{e})}.
    \label{eq:rod-exact-solution}
\end{equation}
Substituting its endpoint derivatives into the junction balances gives a linear system for the unknown vertex temperatures.
Together with the prescribed temperatures, this determines a reference solution independently of the spectral discretization.

\begin{figure}[t]
    \centering
    \includegraphics[width=0.6\textwidth]{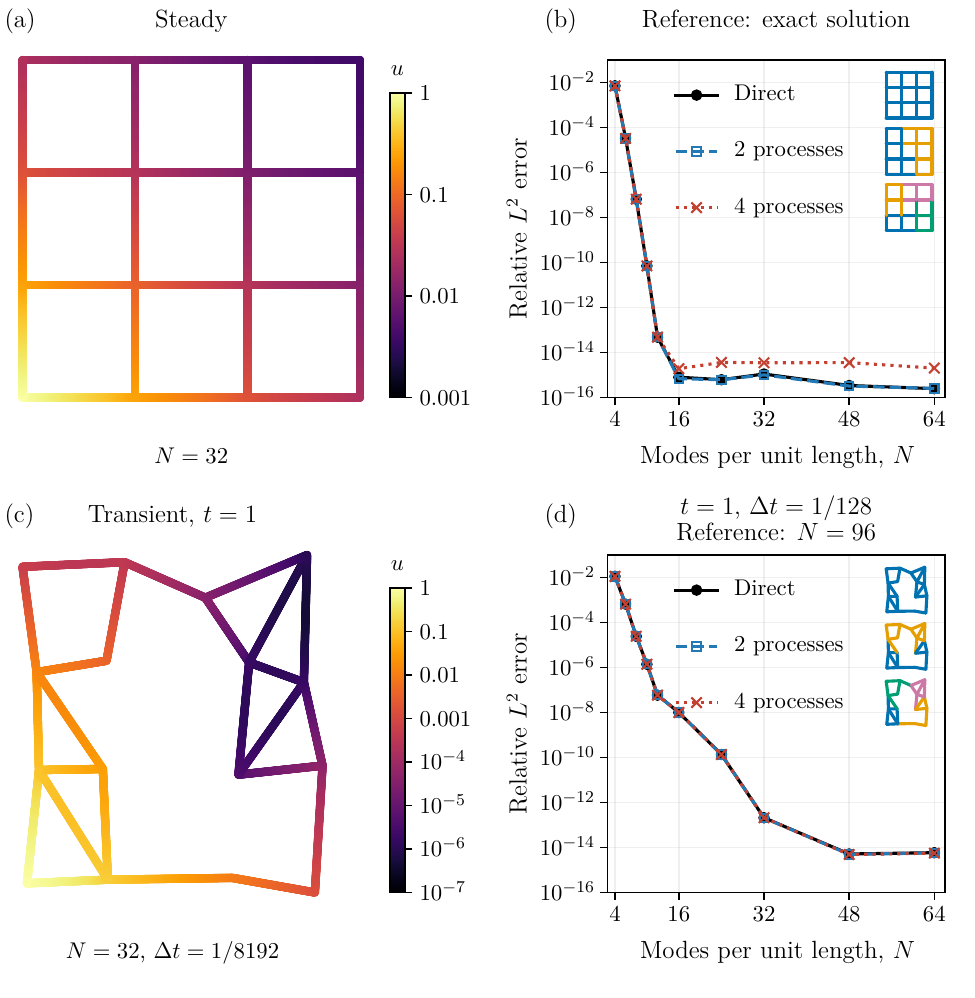}
    \caption{Steady and transient heat conduction in networks with 16 vertices, 24 rods and equal total length.
    (a,b) Steady temperature and relative $L^{2}$ error against the analytical solution for the uniform grid with a heated bottom-left corner.
    (c) Transient temperature at $t=1$ with $\Delta t=1/8192$.
    (d) Relative $L^{2}$ error at $t=1$ against the $N=96$ reference, with $\Delta t=1/128$ throughout the spatial refinement.
    Both temperature fields use $N=32$, giving 768 edge coefficients per network; $N$ is the approximate number of modes per unit length.
    Both colour scales are logarithmic, with different lower limits.
    Convergence curves compare a direct solve with the subgraph method on two and four processes; the miniature networks show the corresponding partitions, coloured by owning process.}
    \label{fig:rod-grid}
\end{figure}

We arrange the rods in a square grid with four vertices in each direction, giving 16 vertices and 24 rods.
In dimensionless variables, each rod has length $\ell_{e}=1$ and coefficients $K_{e}=A_{e}=1$.
We hold the bottom-left vertex at temperature $u=1$ relative to the ambient temperature and impose heat-flux balance at every other vertex.
Heat is conducted along the rods and lost continuously to the surroundings.
The reference solution follows from a linear system for the 15 unknown vertex temperatures.
We vary the number of Chebyshev modes on each rod from $N=4$ to $N=64$ and measure the relative $L^{2}$ error over the whole network, normalized by the reference solution's $L^{2}$ norm.
Figure~\ref{fig:rod-grid}(a,b) shows the temperature decaying away from the heated corner and the spatial convergence of the solution.
The error falls from approximately $7\times10^{-3}$ at $N=4$ to $5\times10^{-14}$ at $N=12$, reaching roundoff levels by $N=16$.
We repeat the calculation on two and four processes, using a spatial partitioner with equal weights for all rods.
The distributed solutions agree with a direct solve of the assembled system to roundoff at every resolution tested.

We next consider transient heating on a heterogeneous network with the same 16 vertices, 24 rods and total rod length of 24.
We perturb the grid vertices, retain 24 edges from their triangulation while preserving connectivity, and rescale the resulting rod lengths to match the grid's total length.
Each rod satisfies
\begin{equation}
    C_{e}\frac{\partial u_{e}}{\partial t}
    -K_{e}\frac{\partial^{2}u_{e}}{\partial x_{e}^{2}}+A_{e}u_{e}=0,
    \label{eq:rod-transient-equation}
\end{equation}
with constant coefficients on each rod, varying across the network within $0.6\leq C_{e}\leq1.4$, $0.7\leq K_{e}\leq1.7$ and $0.15\leq a_{e}\leq0.45$.
Starting at ambient temperature, we prescribe $u=1-\exp(-4t^{2})$ at the bottom-left vertex and impose the same continuity and heat-flux conditions at all other vertices.
We subtract the instantaneous steady temperature profile before time stepping, giving homogeneous vertex conditions and a known forcing term in the edge equations, and restore it when reconstructing the temperature.
We use Dedalus's second-order RK222 scheme and evaluate the solution at $t=1$.

Here $N$ denotes the nominal number of modes per unit length.
We assign integer mode counts approximately in proportion to rod length, with at least four modes per rod and exactly $24N$ edge coefficients in total, matching the uniform grid at each resolution.
The spatial partitioner uses rod lengths as weights.
Figure~\ref{fig:rod-grid}(c) shows the transient temperature for $N=32$ with $\Delta t=1/8192$.
Figure~\ref{fig:rod-grid}(d) shows spatial convergence for $N=4$ to $N=64$ at fixed $\Delta t=1/128$.
We compare each solution with a direct $N=96$ calculation at the same timestep, so these errors measure spatial refinement at fixed $\Delta t$, rather than the total error relative to the continuous-time solution.
The curves reach a roundoff-level plateau near $10^{-14}$; the $N=80$ and $N=96$ references differ by approximately $6\times10^{-15}$ in relative $L^{2}$ norm.
The direct, two-process and four-process curves agree throughout the refinement range.

A separate timestep-refinement study at fixed nominal resolution $N=32$, with $\Delta t$ halved from $1/16$ to $1/1024$, reduces the relative $L^{2}$ error at $t=1$ by approximately a factor of four per refinement, consistent with second-order time integration.
The reference uses $N=40$ and $\Delta t=1/8192$.

\subsection{Benchmark problems and measurement procedure}
\label{sec:benchmark-problems}

We test parallel performance on square grids, branching networks and triangular meshes.
The branching networks contain groups of interconnected vertices, each with a highly connected hub, joined by relatively few edges.
The triangular meshes are periodic square grids with one randomly oriented diagonal added to each cell.

On every network, we evolve two coupled reaction--diffusion fields:
\begin{align}
    C_{u,e}\frac{\partial u_{e}}{\partial t}
    -K_{u,e}\frac{\partial^{2}u_{e}}{\partial x_{e}^{2}}
    +A_{u,e}u_{e}+\frac{1}{5}\, (u_{e}-w_{e})
    & =\frac{1}{10}\,u_{e}^{3}+f_{u,e},
    \label{eq:coupled-benchmark1}
    \\
    C_{w,e}\frac{\partial w_{e}}{\partial t}
    -K_{w,e}\frac{\partial^{2}w_{e}}{\partial x_{e}^{2}}
    +A_{w,e}w_{e}+\frac{1}{5}\,(w_{e}-u_{e})
    &=\frac{1}{10}\, w_{e}^{3}+f_{w,e}.
    \label{eq:coupled-benchmark2}
\end{align}
The coefficients are constant along each edge and vary between edges and fields, with $0.7\leq C\leq1.1$, $0.8\leq K\leq1.1$ and $0.6\leq A\leq0.8$.
Both fields are continuous at vertices and satisfy conductivity-weighted flux balance, without vertex storage or reaction.
The forcing is chosen to provide a known time-dependent solution for checking numerical accuracy.

Edge lengths are sampled uniformly from $[0.5,2]$.
We assign approximately 32 Chebyshev modes per unit length to each field, rounded to multiples of eight, giving 16--64 modes per field on each edge.
Reported coefficient counts include both edge fields and exclude tau variables and vertex unknowns.
We partition the networks using Mt-KaHyPar with coefficient counts as edge weights.

All measurements use AMD Genoa nodes and Dedalus's RK222 scheme with $\Delta t=10^{-3}$.
Each configuration is run once, averaging the wall time over 1,000 timesteps after ten initial steps.
Timings include equation evaluation, linear solves and communication, but exclude network construction, partitioning, assembly, initial factorization, validation and output.

\subsection{Scaling}
\label{sec:scaling}

\begin{figure}[t]
    \centering
    \includegraphics[width=\textwidth]{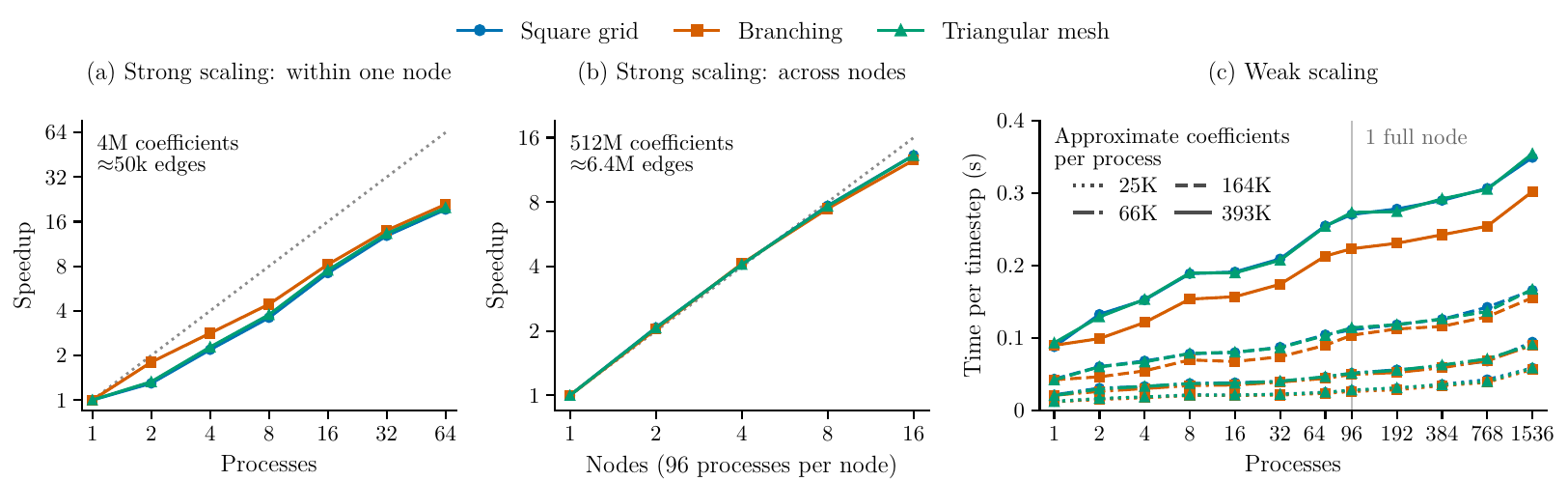}
    \caption{Strong and weak scaling on three network families for equations~\eqref{eq:coupled-benchmark1} and~\eqref{eq:coupled-benchmark2}.
    (a) Within-compute node speedup for approximately four million coefficients, relative to one process.
    (b) Across-node speedup for approximately 512 million coefficients, relative to one node with 96 processes.
    Dotted grey lines indicate ideal speedup.
    (c) Time per timestep at four approximate coefficient counts per process, as indicated in the legend; ideal weak scaling is horizontal.
    The vertical line marks one full node.
    Each point averages 1,000 timesteps.}
    \label{fig:parallel-scaling}
\end{figure}

For strong scaling, we hold the network, coefficients and discretization fixed while increasing the number of processes.
We use approximately four million coefficients (50,000 edges) within one compute node and 512 million coefficients (6.4 million edges) across compute nodes.
Figure~\ref{fig:parallel-scaling}(a,b) shows speedup relative to one process and one full node, respectively.
Within-compute node tests use 1--64 processes; across-node tests use 1--16 nodes with 96 processes per node.

The three networks achieve speedups of 19--21 on 64 processes.
Across compute nodes, scaling is approximately ideal through four nodes, reaching speedups of 12.6--13.2 on sixteen nodes.

For weak scaling, we increase the number of edges while holding the target coefficient count per process fixed.
The edge-length distribution, resolution rule and equations remain unchanged.
The four workloads span approximately 300--5,000 edges per process; small differences from the target sizes arise from integer network dimensions.
Figure~\ref{fig:parallel-scaling}(c) shows the measured time per timestep, which would remain constant under ideal weak scaling.

At the largest workload, timestep time increases by a factor of 2.4--2.9 between one and 64 processes.
Subsequent scaling across fully occupied nodes is substantially better: increasing from one to sixteen compute nodes increases timestep time by 29--35\%, while the total problem size grows approximately sixteenfold.

\section{Examples}

\subsection{EVP: passive optical resonances in photonic networks}

\subsubsection{Model and coupling conditions}

Following \citet{Saxena2022}, we consider passive optical resonances in networks of waveguides constructed from random lines within a square.
Each edge has a constant, possibly complex refractive index $\eta_e$, accounting for absorption.
The degree-one boundary vertices continue into semi-infinite outgoing leads with the same refractive index.

With time dependence $e^{-i\omega t}$, we seek complex resonant wavenumbers $k=\omega/c$, where $c$ is the speed of light, and a continuous electric-field amplitude $u$.
For $k\ne0$, introduce a regular tangent-vector field $q$, with $q|_e=q_e\hat{e}$.
The network equations are
\begin{equation}
    \grad u=ik\,q,
    \qquad
    \div q+\mathcal R u=ik\,\eta^{2}\,u,
    \label{eq:photonic-first-order}
\end{equation}
where
\begin{equation}
    \mathcal R=\sum_{v\in\partial G}\eta_{e(v)}\,\delta_v
    \label{eq:photonic-lead-measure}
\end{equation}
represents the outgoing leads at the open ends $\partial G$, and $e(v)$ is the terminal edge incident to $v$.
The balance equation is evaluated against continuous scalar test fields.

In each edge interior, these equations give the Helmholtz equation
\begin{equation}
    \partial_{x_e}^{2}u_e+\eta_e^2k^2u_e=0.
    \label{eq:photonic-helmholtz}
\end{equation}
At internal vertices, the divergence supplies Kirchhoff coupling,
\begin{equation}
    \sum_{e\sim v}\partial_nu_e(v)=0,
    \label{eq:photonic-kirchhoff}
\end{equation}
where $\partial_n$ is the outward derivative defined in~\eqref{eq:outward-derivative}.
At an open end, the boundary measure instead gives
\begin{equation}
    \partial_nu_e(v)=ik\,\eta_{e} \, u(v),
    \label{eq:photonic-outgoing}
\end{equation}
imposing an outgoing wave in the lead.
The complex resonances account for absorption along the waveguides and radiation through the open ends: $\operatorname{Re}(\omega)$ gives the oscillation frequency and $-\operatorname{Im}(\omega)$ the amplitude decay rate.
The first-order network equations depend linearly on $k$, yielding a generalized eigenvalue problem after spatial discretization.

\subsubsection{Reproduction of the reference result}

We use the network supplied with \citet{Saxena2022}, comprising 96 vertices and 131 edges, with a total internal waveguide length of $2500\,\mu\mathrm{m}$.
The refractive index is $\eta_{e}=1.5$ on terminal edges and their leads, and $\eta_{e}=1.5+0.001i$ on all remaining edges.
We seek passive resonances in the spectral window
\begin{equation}
    10.35\leq\operatorname{Re}(k)\leq11.0,
    \qquad
    0.006\leq-\operatorname{Im}(k)\leq0.015,
    \label{eq:photonic-reference-window}
\end{equation}
with $k$ measured in $\mu\mathrm{m}^{-1}$.
The reference paper reports 454 passive modes in this window.
For the numerical solution, we choose a fixed carrier wavenumber $k_0$ near each spectral target and write
\begin{equation}
    \begin{aligned}
        u_{e}(x_{e})
        &=a_{e}(x_{e})e^{i\eta_{e} k_0x_{e}}
          +b_{e}(x_{e})e^{i\eta_{e} k_0(\ell_{e}-x_{e})},\\
        q_{e}(x_{e})
        &=\eta_{e}\!\left[
          a_{e}(x_{e})e^{i\eta_{e} k_0x_{e}}
          -b_{e}(x_{e})e^{i\eta_{e} k_0(\ell_{e}-x_{e})}
          \right].
    \end{aligned}
    \label{eq:photonic-carrier}
\end{equation}
This exact change of variables gives
\begin{equation}
    a'_{e}=i\eta_{e}(k-k_0)a_{e},
    \qquad
    b'_{e}=-i\eta_{e}(k-k_0)b_{e}.
    \label{eq:photonic-amplitudes}
\end{equation}
We expand the amplitudes in Chebyshev polynomials and impose the vertex conditions on the reconstructed physical fields, obtaining a linear EVP for $k-k_0$.
We use 24 Chebyshev coefficients per amplitude on each edge and repeat the calculation with 32 coefficients to check spatial convergence.

We compare the computed spectrum with a fresh run of the archived netSALT implementation supplied with \citet{Saxena2022}.
For each matched resonance, we compare the complex wavenumber and spatial field profile, accounting for the arbitrary complex amplitude of an eigenmode.
An independent calculation based on exact travelling-wave solutions on each edge provides a further check.
We recover all 454 resonances in the prescribed spectral window, with each mode matched to the archived netSALT rerun and the independent travelling-wave calculation.
Against the latter, the maximum relative differences in $\operatorname{Re}(k)$ and $-\operatorname{Im}(k)$ are below $4\times10^{-16}$ and $3\times10^{-13}$, respectively.
The matched spatial profiles agree to floating-point roundoff in normalized overlap.

\subsubsection{Larger networks}

To assess computational reach, we generate independent networks of the same construction with approximately $10^3$, $10^4$, and $10^5$ physical edges.
We enlarge the physical domain while keeping the local line density, short-distance cutoff, refractive indices and vertex conditions fixed.
At each size, we target 16 resonances near each of six prescribed complex shifts within the reference spectral window, removing duplicates between searches.

\begin{figure}[!t]
    \centering
    \includegraphics[width=\textwidth]{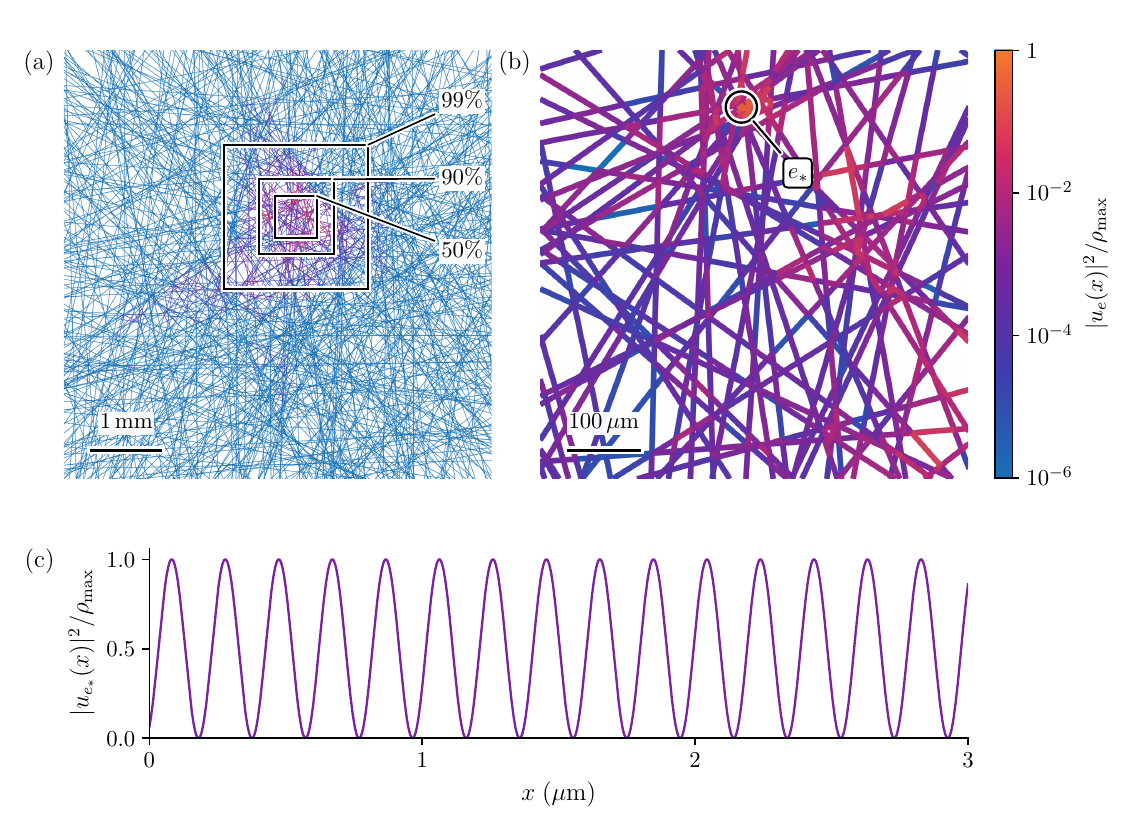}
    \caption{Interior passive resonance of a $100{,}485$-edge network.
    (a) Pointwise intensity $|u_{e}(x)|^{2}$, normalized by its maximum over internal edges, $\rho_{\max}$.
    Nested squares share the internal intensity centroid and contain $50\%$, $90\%$ and $99\%$ of the integrated internal intensity, including partial-edge contributions at their boundaries.
    (b) The $50\%$ square, of width $598.6\,\mu\mathrm{m}$.
    Both maps use the same logarithmic colour scale, with values below $10^{-6}\rho_{\max}$ mapped to its lower endpoint; unresolved optical oscillations are represented by spatially averaged pointwise colours.
    (c) Resolved intensity over the first $3\,\mu\mathrm{m}$ of the circled edge $e_*$, which contains the internal intensity maximum.
    The trace uses the same normalization, with a linear vertical axis and no display floor; its oscillation period is $0.1961\,\mu\mathrm{m}$.
    The resonance has $k=(10.68000379-0.007120003i)\,\mu\mathrm{m}^{-1}$ and is unchanged to numerical precision under refinement from 24 to 32 coefficients per envelope.}
    \label{fig:photonic-interior-mode}
\end{figure}

At the largest size, we compute 96 distinct resonances on each of three networks containing between $100{,}410$ and $100{,}676$ physical edges.
Each mode is checked under spatial refinement and enlargement of the eigensolver subspace.
These are selected resonances near the prescribed spectral targets, rather than complete spectra.
We quantify localization using the integrated intensity
\begin{equation*}
    \sum_{e\ \mathrm{internal}}
    \int_0^{\ell_{e}}|u_{e}(x_{e})|^{2}\,\mathrm{d}x_{e}.
\end{equation*}
Figure~\ref{fig:photonic-interior-mode} shows an interior localized mode on the $100{,}485$-edge network.
Half of this intensity lies within a $0.599\,\mathrm{mm}$ square in the $6.16\,\mathrm{mm}$-wide domain, while the resolved edge profile exhibits an intensity oscillation period of $0.196\,\mu\mathrm{m}$.

\subsection{LBVPs on an evolving network: mineral replacement in porous media}

\subsubsection{Model and coupling conditions}

We use the mineral-replacement model of \citet{BudekSzymczak2025} to demonstrate successive linear boundary-value solves coupled to evolving pore geometry.
Each pore has length $\ell_e(t)$ and uniform diameter $d_e(t)$.
Flow and solute transport are quasisteady relative to mineral evolution.

For a continuous pressure field $p$, the volumetric flow $q$, with $q|_e=q_e\hat{e}$, satisfies
\begin{equation}
    q=-\frac{\pi d^4}{128\mu}\grad p,
    \qquad
    \div q=0,
    \label{eq:mineral-flow}
\end{equation}
where $\mu$ is the fluid viscosity.
The divergence balances are evaluated against continuous test fields that vanish at the inlet and outlet reservoirs.
Their edge and vertex contributions give constant flow along each pore and conservation of flow at internal junctions.
The outlet pressure is zero, and a common inlet pressure is adjusted to maintain total flow $Q$.
Transverse boundaries are periodic.

The reagent and intermediate concentrations $c_B,c_C$ are piecewise scalar fields, using the stoichiometric rescaling of \citet{BudekSzymczak2025}.
Their steady transport satisfies
\begin{equation}
    \begin{pmatrix}
        \div(c_Bq)\\[2pt]
        \div(c_Cq)
    \end{pmatrix}
    =\pi d
    \begin{pmatrix}
        -k_1^{\mathrm{eff}} & 0 \\
        k_1^{\mathrm{eff}} & -k_2^{\mathrm{eff}}
    \end{pmatrix}
    \begin{pmatrix}c_B\\c_C\end{pmatrix}.
    \label{eq:mineral-transport}
\end{equation}
The effective reaction coefficients include diffusion to the pore walls; axial diffusion is neglected.
Dissolution stops when a pore has no remaining access to primary mineral.

The vertex terms conserve each species' advective flux.
We complete the coupling by imposing complete mixing: all outgoing edges receive a common concentration, while incoming concentrations may differ.
Writing $\mathbf c=(c_B,c_C)^{\mathsf T}$, this gives
\begin{equation}
    \mathbf c_e(v)
    =\frac{\displaystyle\sum_{j\in\mathcal I_v}|q_j|\,\mathbf c_j(v)}
           {\displaystyle\sum_{j\in\mathcal I_v}|q_j|},
    \qquad e\in\mathcal O_v,
    \label{eq:mineral-mixing}
\end{equation}
where $\mathcal I_v$ and $\mathcal O_v$ denote incoming and outgoing edges.
At inlet vertices, we prescribe $\mathbf c=(c_{\mathrm{in}},0)^{\mathsf T}$.
Outflow boundaries require no concentration condition.

The uniform pore diameter evolves according to
\begin{align}
    \partial_t d_e
    =\frac{2\Omega_A}{\ell_e}
    \int_0^{\ell_e}
    \left(k_{1,e}^{\mathrm{eff}}c_{B,e}
    -\Gamma k_{2,e}^{\mathrm{eff}}c_{C,e}\right) \mathrm d x_e,
    \label{eq:mineral-diameter}
\end{align}
where $\Omega_A$ is the stoichiometrically weighted molar volume of the primary mineral and $\Gamma$ is the corresponding ratio of precipitated to dissolved solid volumes.
The integrated reaction rates also update the primary and secondary mineral inventories in adjacent grains.
Pore lengths follow the cube-root grain-volume relation of \citet[Eq.~(20)]{BudekSzymczak2025}, and pores below the prescribed closure diameter are removed from the active network.

At each geometry state, Dedalus solves the pressure LBVP and then the coupled transport LBVP with the flow held fixed.
The resulting reaction rates advance the mineral inventories and pore geometry before the next pair of solves.
The complete evolution is nonlinear, while both spatial solves are linear.

\subsubsection{Reproduction of the reference result}

\begin{figure}[t]
    \centering
    \includegraphics[width=\textwidth]{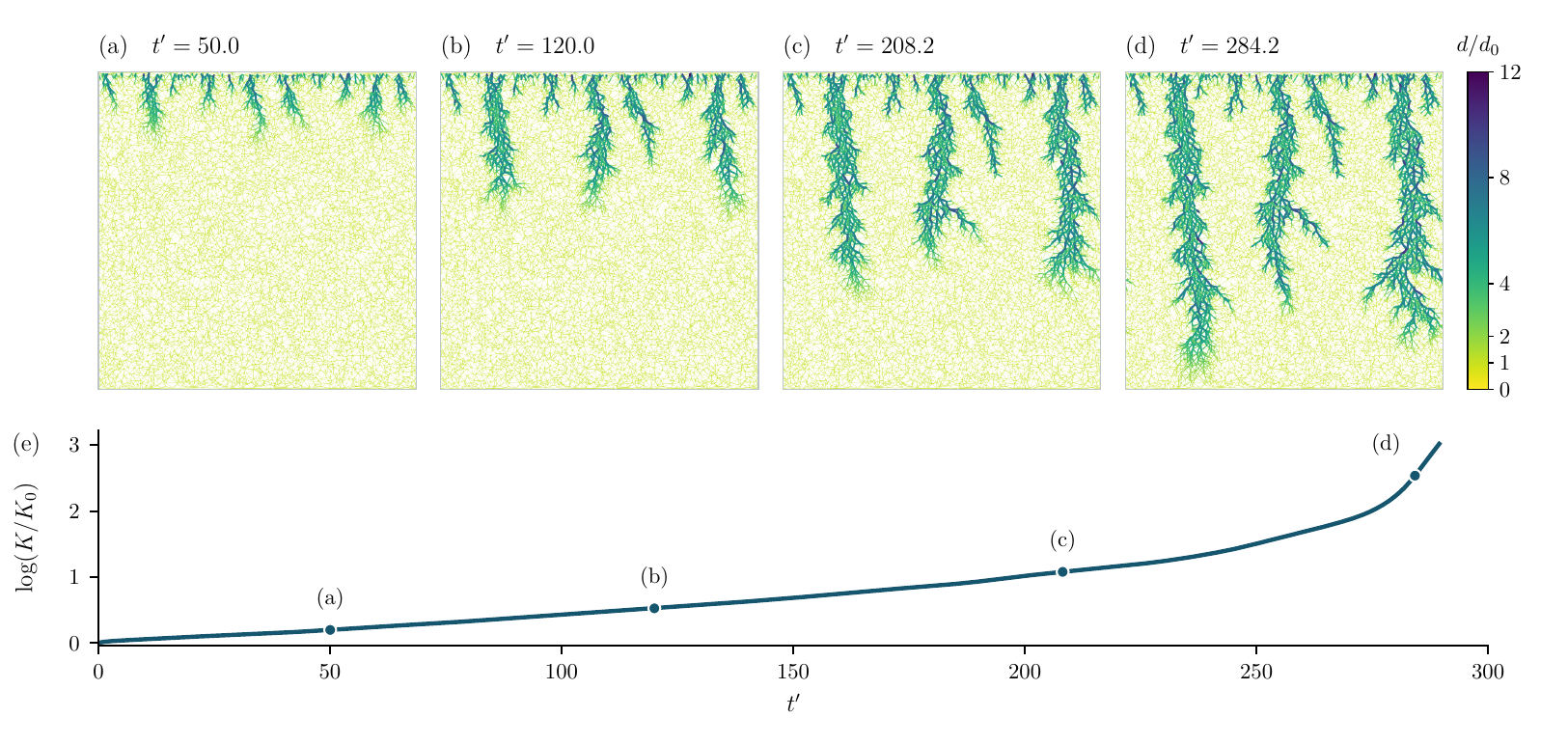}
    \caption{Mineral replacement in the channeling regime ($\Gamma=0.3$, $\kappa=1$, $\mathrm{Da}=0.5$) on an independently generated random network of $29\,996$ pores.
    (a--d) Pore diameters at the indicated times, with flow from top to bottom and periodic transverse boundaries.
    Colours show $d/d_0$ on a shared linear scale, where $d_0$ is the uniform initial diameter; thicker lines indicate wider pores.
    (e) Natural logarithm of permeability relative to its initial value, $\log(K/K_0)$, with the four displayed states marked.
    The dimensionless time is $t'=t/t_0$, where $t_0$ is the dissolution timescale defined by \citet[Eq.~(16)]{BudekSzymczak2025}.}
    \label{fig:mineral-channeling}
\end{figure}

We consider the channeling regime with $\Gamma = 0.3$, $\kappa=1$ and $\mathrm{Da}=0.5$ on an independently generated random network of $29\,996$ pores.
Figure~\ref{fig:mineral-channeling}(a--d) shows channels of enlarged pores extending from the inlet towards the outlet.
Dissolution increases the hydraulic conductance of these paths.
More flow and reagent is directed along these paths, reinforcing the dissolution process.
The channels grow unevenly, with a few extending through most of the network by the final displayed state.
The permeability increases throughout the simulation, with a sharper rise as the channels approach the outlet (Figure~\ref{fig:mineral-channeling}e).
This evolution qualitatively reproduces the preferential channel formation and permeability increase reported by \citet{BudekSzymczak2025}.

\subsection{IVP: nonlinear dispersive waves on a periodic network}
\label{sec:kdv-periodic}

\subsubsection{Model and coupling conditions}

We consider nonlinear dispersive wave propagation on a periodic square network comprising $16\times16$ vertices and $512$ edges of length $\ell_e=8$.
Horizontal and vertical edges are oriented to the right and upwards, respectively, with opposite boundaries identified.
Each vertex therefore connects two incoming and two outgoing edges.

In the network calculus, $\hat{E}$ denotes the prescribed direction field.
For a continuous real scalar field $u$, we define the regular edgewise curvature
\begin{equation}
    w=\hat{E}\dat\grad\!\left(\hat{E}\dat\grad u\right),
    \label{eq:kdv-periodic-network-curvature}
\end{equation}
so that $w_e=\partial_{x_e}^{2}u_e$, without requiring continuity of $w$ at vertices.
The evolution satisfies
\begin{equation}
    \partial_tu+\div\!\left[\hat{E}\left(w+3u^2\right)\right]=0,
    \label{eq:kdv-periodic-network-balance}
\end{equation}
interpreted against continuous scalar test fields.
On each edge, this gives the ordinary nonlinear KdV equation,
\begin{equation}
    \partial_t u_e
    +6u_e\,\partial_{x_e}u_e
    +\partial_{x_e}^{3}u_e=0,
    \qquad 0<x_e<\ell_e,
    \label{eq:kdv-periodic-edge}
\end{equation}
The nonlinear term produces amplitude-dependent propagation, while the third derivative produces dispersion.
At a vertex $v$, label the incoming edges $W,S$ and the outgoing edges $E,N$.
Continuity of $u$ gives
\begin{equation}
    u_W=u_S=u_E=u_N=u(v,t).
    \label{eq:kdv-periodic-continuity}
\end{equation}
The vertex contribution to the divergence balances the incoming and outgoing fluxes.
Because $u$ is continuous and the crossing is balanced, the $3u^2$ contributions cancel, leaving
\begin{equation}
    w_W+w_S=w_E+w_N.
    \label{eq:kdv-periodic-curvature}
\end{equation}
We complete the coupling by prescribing derivative conditions
\begin{equation}
    u'_W=u'_E,\qquad u'_S=u'_N.
    \label{eq:kdv-periodic-slopes}
\end{equation}
Together, these are the balanced $k=m=2$, $B=I$ conditions considered by \citet[Example~8.3]{CavalcanteMarques2025}.
For smooth solutions, testing the network balance with unity gives conservation of mass.
Testing with $u$ and using the slope pairings gives conservation of the squared $L^2$ norm:
\begin{gather}
    \frac{\mathrm d}{\mathrm dt}
    \sum_{e\in E}\int_0^{\ell_e}u_e\,\mathrm dx_e=0, \quad
    \frac{\mathrm d}{\mathrm dt}
    \sum_{e\in E}\int_0^{\ell_e}u_e^2\,\mathrm dx_e=0.
\end{gather}
We monitor deviations from their initial values to assess numerical conservation.

\subsubsection{Pulse evolution}

We initialize a localized pulse on one horizontal edge $e_*$,
\begin{equation}
    u_{e_*}(x,0)
    =\left[1-\left(\frac{x-4}{4}\right)^2\right]^8,
    \label{eq:kdv-periodic-initial}
\end{equation}
and set $u_e(x,0)=0$ elsewhere.
We use a Legendre basis with $64$ modes per edge. 
Junction conditions are enforced using correction functions constructed from their adjoints with respect to the physical \(L^2\) inner product.
Time integration uses the RKGFY scheme with a timestep of $1.25\times10^{-4}$, and nonlinear terms are evaluated with $3/2$ dealiasing. 
Resolution is checked by spatial and temporal refinement.
Over $0\leq t\leq200$, the maximum relative drifts in total mass and squared $L^2$ norm are $6.00\times10^{-10}$ and $2.07\times10^{-8}$, respectively.

\begin{figure}[H]
    \centering
    \includegraphics[width=0.8\textwidth]{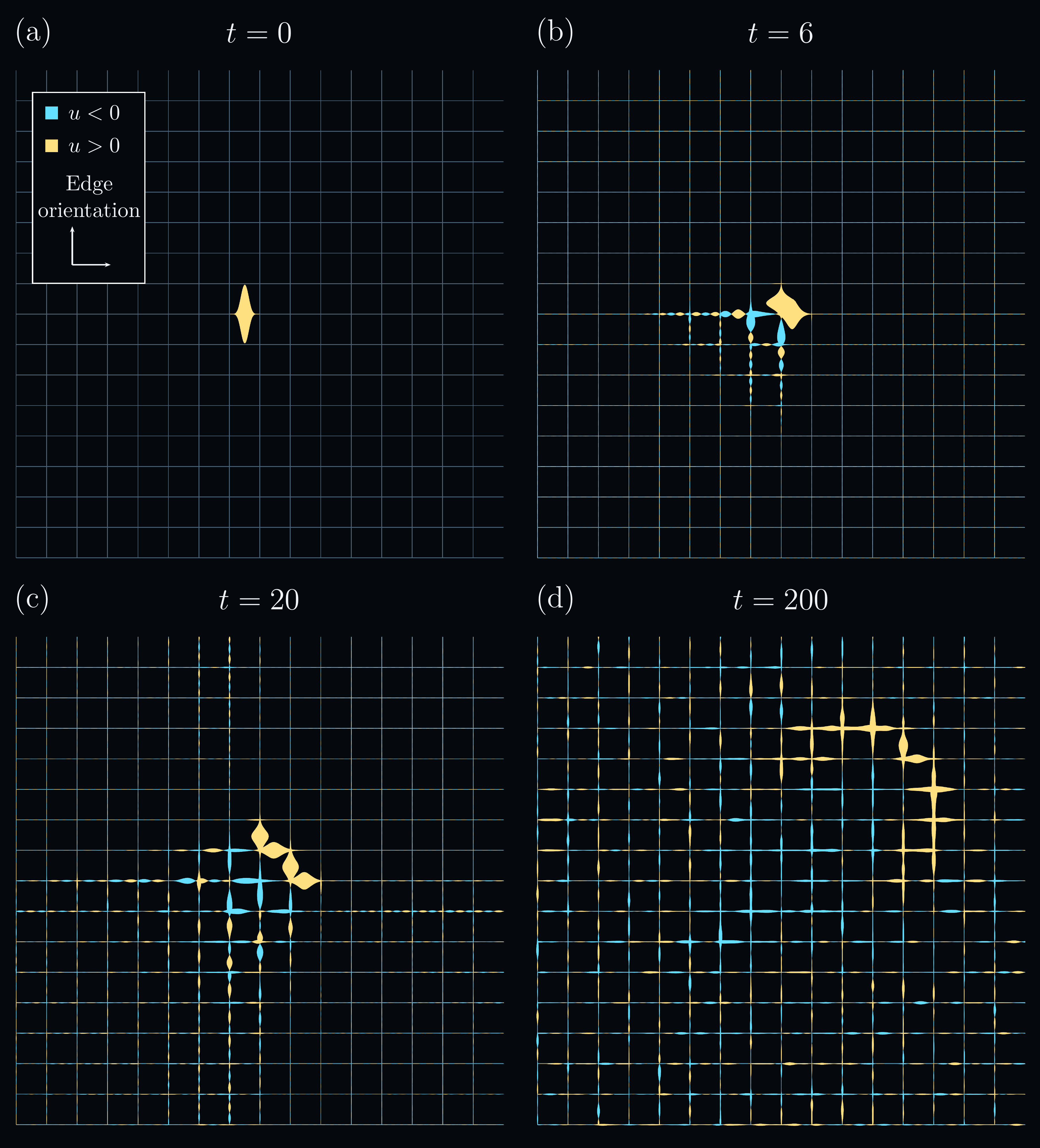}
    \caption{KdV pulse evolution on a $16\times16$ periodic network.
    Gold and blue indicate positive and negative $u$, respectively; pulse width is proportional to $|u|$.
    Thin lines show the network geometry. 
    The edge orientation field, $\hat{E}$, is rightward on horizontal edges and upward on vertical edges.
    The outgoing endpoints at the right and top boundaries are identified with the corresponding vertices on the left and bottom boundaries.}
    \label{fig:kdv-pulse}
\end{figure}

Figure~\ref{fig:kdv-pulse} shows the evolution at $t=0,6,20,200$.
The initially nonnegative pulse develops oscillatory profiles of both signs and transfers amplitude between horizontal and vertical edges.
Although the initial disturbance occupies a single horizontal edge, vertical edges carry approximately $52\%$ of the total squared $L^2$ norm by $t=6$.
By $t=200$, the positive peak has fallen from $1$ to approximately $0.157$.
The conservation measurements show that this decrease accompanies redistribution across the network with negligible loss of total squared amplitude.

To distinguish the effect of nonlinearity, we compare with a linear control obtained by omitting $6u_e\partial_{x_e}u_e$, retaining the same initial pulse and coupling conditions.
We measure concentration across edges by
\begin{equation}
    Q_e(t)=\int_e u_e(x,t)^2\,\mathrm dx,
    \qquad
    N_{\mathrm{eff}}(t)
    =\frac{\left(\sum_e Q_e(t)\right)^2}
    {\sum_e Q_e(t)^2}.
    \label{eq:kdv-periodic-participation}
\end{equation}
This quantity equals one when the squared amplitude is confined to one edge and equals $m$ when it is distributed equally among $m$ edges.
At $t=200$, $N_{\mathrm{eff}}$ is approximately $89$ for the nonlinear solution and $203$ for the linear control.
The nonlinear solution retains a more concentrated distribution across edges at this time.

\section{Discussion}

We have developed a concise yet general numerical framework for solving PDEs on physical networks. 
The contributions include a robust conceptual framework for defining edge differential equations and consistent coupling conditions at vertices, as well as a high-performance distributed spectral solver for simulation.

The network calculus allows differential equations to be written in terms of fields over the whole network, with edge and vertex equations appearing in the same expressions.
The practical value is that relations between these contributions remain explicit as a model is developed.
Integration by parts relates differential operators to conservation laws, and field evaluations at vertices and their adjoints connect kinematic constraints to reaction forces.
The beam example requires a choice of joint kinematics, the KdV example requires directional data and slope pairings, and the mineral model assumes complete mixing in addition to species-flux conservation.

This formulation builds upon Dedalus’s existing field and equation interface.
Cardinal bases collect the edge expansions into network fields, while the connectivity maps supply the relations between edge and vertex data. Polynomial differentiation, nonlinear evaluation, time integration and eigenvalue algorithms reuse the existing Dedalus machinery.

The example problems of the paper demonstrate the different capabilities of the infrastructure at a variety of scales.
The photonic EVP reproduces the reference spectrum and then computes selected resonances on networks with approximately $10^5$ edges.
The mineral example couples pressure and transport LBVP solves to pore evolution, producing channels and increased permeability.
The KdV IVP calculation uses the full functionality of the network calculus.
From an initial localized pulse, the solution preferentially splits in the northward and eastward directions selected by the edge orientations.
Through periodicity of the square grid, backward-propagating dispersive radiation interacts with the positive-amplitude pulse fragments as the solution spreads.
The long-time dynamics remain unresolved: the evolution could approach a broadly distributed dispersive state or exhibit pulse recombination and recurrence.

Finally, we envisage ongoing maintenance and improvements as applications require.
Adaptive spectral resolution could concentrate computational effort on edges where finer structure develops--this could be particularly useful on hierarchical or multiscale networks--while stronger interface preconditioners could improve parallel scaling.
For evolving networks, adapting the distribution of work as the geometry changes may become worthwhile.

\subsection*{Code and data availability}

The source code, example scripts, and data needed to reproduce the figures are available at \url{https://github.com/sidneyholden1/dedalus-networks}.

\subsection*{Declaration of AI use}

The authors conceived the study, devised the overall algorithms, designed the Dedalus computational interfaces, and selected the examples and drafted the manuscript. OpenAI’s ChatGPT and Codex were used to assist with literature searches and identification of references, implementation of numerical routines, collection of simulation data, preparation of figures, and revision of the manuscript. 
The authors retain full responsibility for the mathematical and computational results, their interpretation, and the content of the manuscript.

\bibliographystyle{unsrtnat}
\bibliography{references}

\end{document}